\documentclass[11pt,a4paper]{article}
\usepackage[margin=2cm]{geometry}
\usepackage{graphicx}
\usepackage{caption}
\usepackage{subcaption}
\usepackage{amsmath}
\usepackage{amssymb}
\usepackage{float}

\title{Vertical Structure-Based Composites for Analyzing Atmospheric Front Environments}
\author{Anne Gossing, Miriam Kliemann, Vijay Natarajan, Stephan Pfahl, Daniel Baum}

\begin{document}

\maketitle

\begin{abstract}
In many mid-latitude regions, a substantial fraction of extreme precipitation events occur in the vicinity of atmospheric fronts. 
To understand the processes that lead to such extreme precipitation, it is important to study the environment of the front. 
In this paper, we introduce a new method for computing vertical composites of frontal environments. 
Existing approaches construct such composites relative to front lines that represent the front at a single pressure level. 
However, fronts are vertically tilted and can exhibit substantial geometric variability, both within an individual front and between distinct events.
Therefore, at levels away from the reference front line, the alignment of front environments based on a single front line commonly blurs the signals in the analyzed atmospheric variables. 
Our method accounts for the vertical structure of the front by computing composites relative to the actual front position at all height levels. 
We showcase the method using ERA5 reanalysis data for 50 cold fronts over Europe and apply it to thermodynamic and kinematic fields. 
Compared with conventional line-based composites, the new approach yields signals that are sharper and spatially more coherent. 
It also provides a visualization in which front-relative distances can be interpreted directly throughout the vertical column. 
Our method offers a general framework to study atmospheric processes in frontal environments and can be applied to different types and definitions of fronts.
\end{abstract}

\section{Introduction}
\label{sec:intro}

Atmospheric fronts are boundaries between air masses with different properties, typically different densities, and thus different temperatures.
Such fronts are a central component of mid-latitude weather variability.
For example, a substantial fraction of mid-latitude extreme precipitation occurs in the vicinity of fronts~\cite{cattoPrecipitationFronts2012,cattoImportanceFrontsExtreme2013} and fronts also shape the projected intensification of extreme precipitation in a warmer future climate~\cite{konstaliFrontsFuturePrecipitation2025}.
In the warm season, the environmental conditions during the passage of cold fronts can be favorable for the occurrence of deep convection, which not only causes precipitation, but potentially also other hazards such as lightning and strong winds~\cite{paceyClimatologyNatureWarmseason2023,paceyEnvironmentsLiftingMechanisms2025}.
Furthermore, fronts are embedded in larger-scale circulation anomalies, and the mutual interactions between fronts and mid-latitude cyclones, e.g., via frontal wave cyclogenesis, can have a strong influence on the development of these weather features~\cite{shapiroFronts1990,hewsonObjectiveIdentificationFrontal1997,schemmWhenTheirLife2018,lichteneggerColdFrontalLife2025}.

To study these various processes and impacts associated with fronts and quantify their relevance beyond individual case studies, methods for the automatic detection of fronts have been developed since the 1960s~\cite{hewsonObjectiveFronts1998,simmondsIdentificationClimatologySouthern2012,renardNumericalFrontalAnalysis2015,thomasGlobalClimatologiesFronts2019,sansomIdentificationClimatologies2024}.
More recently, machine-learning methods have been developed for the automated detection and classification of frontal features~\cite{lagerquist2020ClimatologyVariabilityWarm,nieblerAutomatedDetectionClassification2022,dagon2022MachineLearningBasedDetection, justin2025FrontFinderEfficient}.
The available methods differ, e.g., with regard to the atmospheric variables on which the detection is based.
Such variables include potential temperature~($\theta$), which is the temperature that an air parcel would have if brought adiabatically to a reference pressure of 1000\ mb, equivalent potential temperature~($\theta_e$), which additionally accounts for latent heat release during cloud formation, and horizontal wind.
These methods also differ in the mathematical function that defines the frontal boundary, such as the magnitude of a horizontal gradient or the thermal front parameter~(TFP), which is the negative directional derivative of the horizontal gradient magnitude in the direction of the horizontal gradient~\cite{Thomas2019-hu}.
For most of these approaches, a specific two-dimensional domain (often a level of constant pressure) is selected on which the front is identified in the form of a line.
Pressure levels are more suitable for this purpose than, for example, altitude levels because the relationship between pressure, temperature, and density can lead to comparable thermodynamic states being located at different geometric heights in the atmosphere.
However, fronts also possess a vertical structure, including a vertical slope of the boundary between cold and warm air~\cite{martinFronts2024}.
For a comprehensive analysis and understanding of the complex dynamics of fronts and their relation to impacts, variations in the vertical structure, which are not taken into account in such classical two-dimensional automatic detection methods, can play an important role.
For example, vertical differences between the relative motion of warm and cold air are associated with structurally different cold frontal types, also denoted as ana- and kata-fronts~\cite{bergeron1937physics, sansomIdentificationClimatologies2024}, which go along with differences in the spatial distribution of clouds, convection, and precipitation relative to the front~\cite{browningSynopticColdFronts1982}.
The limitation caused by looking at a single front line instead of the entire vertical structure of the front can be mitigated by computing front lines at several two-dimensional domains and stacking these front lines together to create a front surface.
Niebler et al.~\cite{niebler_scalable_2024}, e.g., apply such an approach.
However, it is not guaranteed that these front lines really describe a contiguous front surface since the front lines are extracted independently in a slice-by-slice fashion, that is, in 2D rather than by working directly in 3D.
Hence, recent studies have begun to develop three-dimensional detection methods for the identification of fronts as two-dimensional surfaces separating two air masses~\cite{kernInteractive3DVisual2019,beckertThreedimensionalStructureFronts2023,gossing2024ridge}.
Beckert et al.~\cite{beckertThreedimensionalStructureFronts2023}, following Kern et al.~\cite{kernInteractive3DVisual2019}, use a contour-based method on smoothed data from numerical weather prediction models to investigate details of frontal dynamics, such as the temporal evolution of the front or its interaction with specific air streams, such as the warm conveyor belt.
In contrast, Gossing et al.~\cite{gossing2024ridge} use ridge surface computation, which among other properties differs from the approach by Kern and Beckert et al.\ by only requiring the computation of lower-order derivatives of the underlying atmospheric field, which is numerically more robust and therefore does not require data smoothing.

Based on one-dimensional front lines extracted with two-dimensional front detection algorithms, Pacey et al.~\cite{paceyClimatologyNatureWarmseason2023} and Schaffer et al.~\cite{schafferDriversColdFrontalPrecipitation2024} recently investigated the mean environment of European cold fronts using vertical cross-frontal composites.

However, their method uses only a single point in each vertical cross-frontal slice, that is the intersection point of the one-dimensional front line at a certain pressure level with the vertical cross-section, to apply a global alignment (translation) of the entire vertical cross-section.
This leads to a substantial smoothing of the atmospheric quantities in the near horizontal vicinity of the front for all pressure levels other than the one at which the front line is identified.

In this paper, we address this limitation and extend their approach by using the entire two-dimensional front surface for computing the vertical cross-frontal composites.
Instead of aligning the vertical cross-frontal slices with respect to only a single point on each slice, our proposed method aligns each row of the vertical cross-section separately.
As we will show below, this results in more pronounced structural features of the atmospheric variables in the computed vertical cross-frontal composites that to a large extent are smoothed out in the previous approach.
We propose two visualizations of the vertical cross-frontal composites.
First, we use a completely straight vertical front line as reference. 
Second, we align the vertical cross-frontal slices to the average vertical front line.
Whilst the first visualization does not represent a natural scenario, because it shows the front as a straight upright structure, it does allow for an easy visual readout of the atmospheric variables w.r.t.\ the distance to the front at all pressure levels.
In comparison, the second visualization depicts the same values as the first one but in a different, more natural and intuitive way.

For the computation of the two-dimensional front surfaces, we apply the algorithm of Gossing et al.~\cite{gossing2024ridge}.
All front surfaces are extracted as contiguous two-dimensional ridge surfaces from three-dimensional reanalysis data.
To demonstrate the benefits of our new method for computing vertical composites, we use a set of 50 synoptic-scale cold fronts of at least 1000\, km length in Central Europe during the warm season.
For these 50 cold fronts, we compute cross-frontal vertical composites of thermodynamic and kinematic fields with the previous approach as well as the new approaches, and compare the results.

In order to facilitate an easy reading of the manuscript, key mathematical notations that are introduced within individual sections are summarized in Table~\ref{tab:notations} within Appendix~\ref{sec:notations}.

\section{Data}
\label{sec:data}

In this study, we use ERA5 global reanalysis data~\cite{hersbachERA5GlobalReanalysis2020},
which represent three-dimensional atmospheric data given on a regular grid with a spacing of 0.25° in both latitude and longitude.
In particular, we use ERA5 data for the domain $30^{\circ}$\,N~--~$70^{\circ}$\,N and $20^{\circ}$\,W~--~$30^{\circ}$\,E, for the period 2007~--~2016, restricted to the months May\,--\,September, sampled at 12\,UTC (Coordinated Universal Time).
Vertically, 23 pressure levels between 200\,hPa and 1000\,hPa are linearly interpolated to obtain a regular grid spacing of 10\,hPa.
The data used are summarized in Table~\ref{tab:list_fronts} in Appendix~\ref{sec:fronts}.

For the computation of composites of vertical cross-sections and to support distance analysis, the data is further transformed from the latitude-longitude grid to a horizontally curvilinear grid with metric coordinates $(X,Y)$, while the vertical coordinates and resolution are not modified.
The horizontal transform is computed relative to a reference point $(\lambda_0,\varphi_0)$, chosen as the center of the analysis domain, using a spherical Earth approximation with $R=6371$\,km:
\begin{align}
Y &= R \, (\varphi-\varphi_0),\\
X &= \,R \, \cos(\varphi)\,(\lambda-\lambda_0),
\end{align}
with angles in radians.

To derive the thermal quantity called the equivalent potential temperature, denoted by $\theta$$_e$, we use temperature, specific humidity, and pressure.
The variable $\theta$$_e$ is conserved during vertical motion, including latent heat release associated with moist processes in the atmosphere.
Fronts identified using $\theta_e$ are characterized by strong gradients of temperature, humidity, or both~\cite{thomasGlobalClimatologiesFronts2019}.
However, temperature and humidity gradients are not necessarily colocated, and strong $\theta_e$ gradients do not always represent synoptic fronts~\cite{Thomas2019-hu}.
Therefore, in this work, $\theta_e$ is used to extract continuous front features associated with previously identified fronts.

\section{Methods}
\label{sec:methods}

\subsection{Objective Front Definition}
\label{sec:methods:frontdefinition}

As discussed in Section~\ref{sec:intro}, previous studies used various atmospheric fields as a basis for automatic detection of fronts.
Here, to allow for a comparison with previous work and consider fronts that have been shown to be relevant for the occurrence of convection over Europe in the summer season, we use the front dataset of Pacey et al.~\cite{paceyClimatologyNatureWarmseason2023}.
In this dataset, fronts were objectively detected following Hewson et al.~\cite{hewsonObjectiveFronts1998} with the equivalent potential temperature as thermodynamic input variable.
We use these previously identified fronts to extract the corresponding two-dimensional front surfaces from three-dimensional fields of $\theta_e$.
However, unlike Pacey et al., we do not smooth the data prior to further processing.
For the ridge detection described below, we calculate the magnitude of the first horizontal derivative $\lVert\nabla_h\theta_e\rVert$ of this (unsmoothed) field~\cite{gossing2024ridge}.
The directional maxima of the first derivative can be used to identify the surface with the largest horizontal gradient of $\theta_e$ in the frontal zone, which serves as the main characteristic of the front.
This is consistent with the front definition of Pacey et al.~\cite{paceyClimatologyNatureWarmseason2023}, since these maxima coincide with the zero contours of the thermal front parameter~(TFP) used to define their one-dimensional front lines.
This way, fronts are represented by a two-dimensional surface along the center of the frontal zone. 
Although fronts are conventionally placed at the warm-air boundary of the frontal zone, we use this representation to remain consistent with the front dataset of Pacey et al.~\cite{paceyClimatologyNatureWarmseason2023}. 
Alternatively, the warm-air boundary could be extracted as a surfaces from the ridges of the negative TFP field. 
The 2D front surface extraction method used in this work is described in the following section.

\subsection{2D Front Surface Extraction}
\label{sec:methods:ridgesurfaceextraction}

\begin{figure}[h]
    \centering

    \begin{subfigure}[b]{0.96\linewidth}
        \centering

        \fbox{\includegraphics[width=0.96\linewidth, trim=0.15cm 1.1cm 0.72cm 0.32cm, clip]{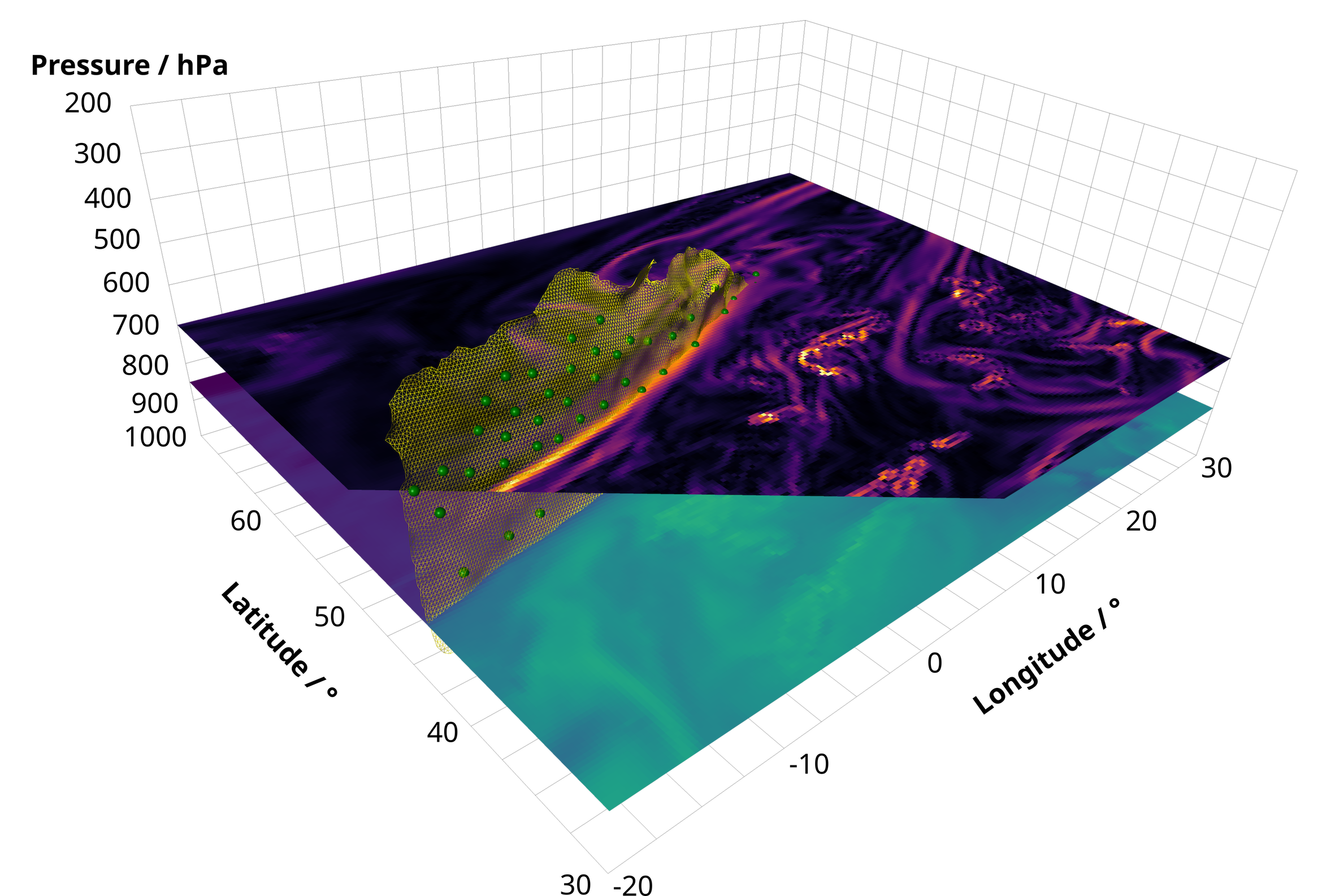}} 
    \end{subfigure}

    \vspace{0.3cm}

    \begin{subfigure}[b]{0.89\linewidth}
        \centering
    \includegraphics[width=0.49\linewidth]{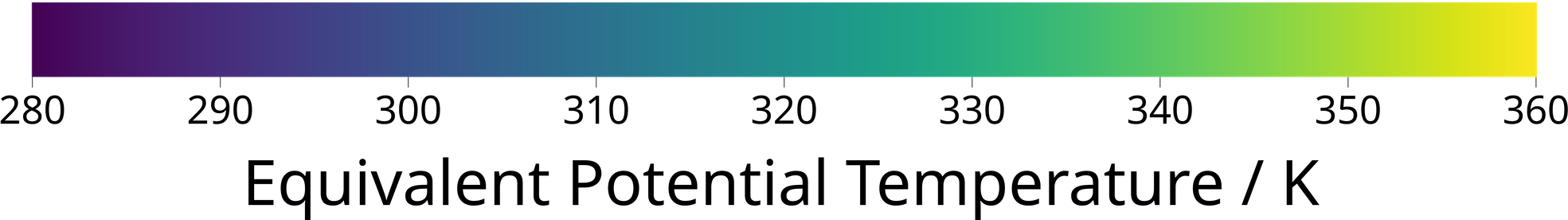}
    \hspace{0.04cm}
    \includegraphics[width=0.49\linewidth]{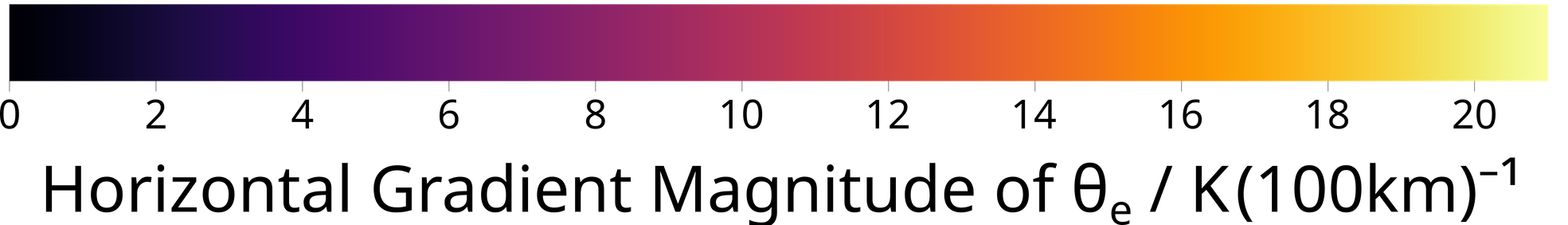}

        \end{subfigure}
    \caption{Ridge surface (yellow triangular mesh) extracted from the 3D scalar field of the horizontal gradient magnitude of $\theta$$_e$ (shown as a horizontal slice at 700\,hPa). The underlying equivalent potential temperature field $\theta$$_e$ is shown at 850\,hPa. The green points depict the seed points used for computing the ridge surface.}
    \label{fig:surface_extraction}
\end{figure}

In this section, we describe the method that we used to extract the 2D front surfaces.
Since 2D front surfaces are an essential part of the proposed workflow, we describe the method to ensure the completeness of the manuscript.
While a specific front surface extraction method is employed to present our results, it is important to note that our proposed workflow for computing vertical composites supports other methods that compute contiguous 2D front surfaces.
Indeed, small variations in outcome are anticipated due to the inherent differences between the front extraction methodologies.

In this work, as previously proposed by Gossing et al.~\cite{gossing2024ridge}, we extract the 2D front surfaces of cold fronts using the ridge surface extraction method developed by Klenert et al.~\cite{klenertLocalIterativeApproach2023}.
This method uses a fast-marching scheme~\cite{sethianFastMarchingMethods1999b,algarniSurfCutSurfacesMinimal2017} to locally compute surface patches that are subsequently merged into a contiguous surface mesh~\cite{klenertLocalIterativeApproach2023}.
The ridge extraction method takes as input two types of data: (1)~a three-dimensional (3D) scalar field containing two-dimensional (2D) ridge structures representing the cold fronts, and (2)~a set of seed points for the local fast-marching algorithm, where each seed point generates a single surface patch of the cold front.
For cold fronts, the horizontal gradient magnitude of the equivalent potential temperature field (cf.~Sect.~\ref{sec:data}) can be used as a scalar field containing the 2D ridge structures (see~Fig.~\ref{fig:surface_extraction}).
The resulting 2D front surface is represented by a triangular mesh that follows the highest values of the magnitude of the horizontal gradient of the equivalent potential temperature.
In other words, the 2D mesh separates two regions from each other, one with rather cold and dry air from that of warm and humid air.
This is depicted in Figure~\ref{fig:surface_extraction} by showing the extracted surface with both the equivalent potential temperature field and its gradient magnitude.
One can clearly observe that the surface mesh is located at the positions where the gradient magnitude field has its locally highest values.
The seed points needed for the surface extraction can be generated interactively or automatically, where one assumption about the seed points is that they lie on or close to the 2D ridge structure of interest.
The ridge surface extraction method naturally stops in regions with small scalar values, since for small values the fast-marching costs become very large.
This leads to ridge surfaces that vanish at low gradients of the equivalent potential temperature field, which is sensible as a low gradient indicates that either there is {no ridge structure anymore or the distinction between the air masses w.r.t.\ temperature and humidity is very low (cf.~Sect.~\ref{sec:methods:frontdefinition}).
In both cases, ridge surface extraction should stop.
The extracted 2D ridge surface is represented as a triangular surface mesh where each vertex lies at the local directional maximum of the gradient field of the equivalent potential temperature defined by the normal of the surface at the respective vertex.

To generate seed points for the ridge surface extraction method, we initially tried using the front lines generated by the algorithm of Pacey et al.~\cite{paceyClimatologyNatureWarmseason2023}.
These front lines were sampled with regular spacing, resulting in a set of seed points.
However, since Pacey et al.\ smoothed the equivalent potential temperature field before computing the gradient, the seed points generated often did not coincide with the ridge structure of the unsmoothed data, depending on the horizontal curvature of the front.
Although the ridge surface extraction algorithm is quite robust to the placement of the seed points, they were too far away from the front that we wanted to extract in regions with high curvature or where multiple small-scale front features were merged into one, due to smoothing.
This problem could be partially circumvented by moving the initial seed points along the stream lines of the gradient field to their local maxima along the stream lines.
However, this also did not always give satisfactory results, since when the misplacement was too large, it could happen that the seed points were moved to the wrong neighboring ridge structure.
Therefore, instead of using all points along the front line, we only used a few manually selected points as initial seeds for which front surface patches were generated.
From these initial seed points, we then used the automatic seeding method implemented in the ridge surface extraction method~\cite{klenertLocalIterativeApproach2023}.
After generating the surface patches for the initial seed points, the method automatically places further seed points at the borders of the current front surface, as long as the new seed points enlarge the front surface further.
This worked sufficiently well for most of the front cases considered here.
In some cases, additional seed points were placed interactively to further increase the front surface in regions where the gradient of the equivalent potential temperature was weak.
Further automation of seed point placement will be subject to future work and is discussed in Section~\ref{sec:discussion}.

\subsection{Vertical Frontal Composite Computation}
\label{sec:methods:frontalenvanalysis}

In this section, we detail our novel approach to computing vertical frontal composites.
The basis for this approach is 2D front surfaces extracted using the method described in Section~\ref{sec:methods:ridgesurfaceextraction}.
These front surfaces are represented as 2D triangular meshes that separate cold and dry air from warm and humid air.
Assuming that we have already computed the 2D front surfaces for all fronts of interest, our approach starts by extracting the horizontal front lines of the 2D front surfaces at some pressure level $\mathcal{L}_{\text{p}_0}$ and representing the front lines as a parametric curve with an orientation associated with each point of the curve (Sect.~\ref{sec:methods:frontalenvanalysis:horizontalfrontline}).
At regular sample points of the horizontal front line, we then extract vertical slices from the 3D volumes (Sect.~\ref{sec:methods:frontalenvanalysis:sampling}).
Additionally, at the same sample points of the horizontal front line, we compute the vertical front line by intersecting the 2D front surface with the respective vertical slice computed previously (Sect.~\ref{sec:methods:frontalenvanalysis:verticalfrontline}).
One of the proposed visualizations requires the calculation of the mean vertical front line, which is described in Section~\ref{sec:methods:frontalenvanalysis:verticalfrontlinemeans}.
Finally, the vertical slices are averaged in two different ways, resulting in different types of vertical composites (Sect.~\ref{sec:methods:frontalenvanalysis:mean}).

\subsubsection{Computation of Horizontal Front Line}
\label{sec:methods:frontalenvanalysis:horizontalfrontline}

Let $\mathcal{F}_\text{S}$ denote a front surface extracted using the ridge surface extraction approach described in Section~\ref{sec:methods:ridgesurfaceextraction}.
Furthermore, let $\mathcal{P}_{\text{H}\mathcal{L}_{\text{p}_0}}$ denote the horizontal plane at the pressure level $\mathcal{L}_{\text{p}_0}$.
We define the horizontal front line $\mathcal{F}_{\text{HL}}$ as the intersection $\mathcal{F}_{\text{HL}} := \mathcal{F}_\text{S} \cap \mathcal{P}_{\text{H}\mathcal{L}_{\text{p}_0}}$, and represent it as a polyline, sampled with a regular spacing $\mathcal{S}_\text{HL}$.
This yields a set of points along the horizontal front line, where we denote its $i$th point by $\mathcal{F}_{\text{HL}}(i)$.
With each sample point $\mathcal{F}_{\text{HL}}(i)$ of the line, we associate an orientation represented as the normal direction $\mathcal{N}_{\text{HL}}(i)$ at this point, and compute it as the normal of the line connecting its predecessor $\mathcal{F}_{\text{HL}}(i-1)$ and successor $\mathcal{F}_{\text{HL}}(i+1)$, lying in the plane $\mathcal{P}_{\text{H}\mathcal{L}_{\text{p}_0}}$.
For consistency, we ensure that all normals point to the same side of the line.
For cold fronts, the normals point to the warm-side of the front.

\subsubsection{Sampling of Vertical Slices}
\label{sec:methods:frontalenvanalysis:sampling}

For the analysis of the frontal environment, the thermodynamic and kinematic variables are evaluated in vertical slices.
The horizontal orientation of these vertical slices is determined from a horizontal front line (Sect.~\ref{sec:methods:frontalenvanalysis:horizontalfrontline}) given at some pressure level $\mathcal{L}_{\text{p}_0}$.
Pacey et al.~\cite{paceyClimatologyNatureWarmseason2023} used $\mathcal{L}_{\text{p}_0} = 700\,hPa$ because it reduces interference from orography and boundary-layer turbulence, producing smoother and more reliable automatically detected frontal lines.
Our method is not restricted to a specific pressure level, so any level at which the front is clearly visible could be used.
However, when using vertical composites for statistical analyses or to compare different cases, the same pressure level should be used.
For each point $\mathcal{F}_{\text{HL}}(i)$ of the horizontal front line, we define a vertical slice as the plane spanned by two vectors $\mathbf{u}_i$ and $\mathbf{v}$, where $\mathbf{u}_i = \mathcal{N}_{\text{HL}}(i)$ is given by the normal vector of the horizontal front line at point $\mathcal{F}_{\text{HL}}(i)$, and $\mathbf{v}=\mathbf{e}_z=(0,0,1)$ is parallel to the global vertical axis.
Within each slice, the horizontal coordinate is defined relative to the horizontal front line point, such that $u=0\,km$ corresponds to $\mathcal{F}_{\text{HL}}(i)$.
In this vertical slice, we now sample the field of interest on a regular grid G (see~Fig.~\ref{fig:spine_and_grid}) with user-defined horizontal extent $\mathcal{E}_\text{HG}$, e.g., $\pm 500$\,km, and horizontal spacing $\mathcal{S}_\text{HG}$, e.g., of 10\,km, while the vertical extent of the slice follows the entire vertical range of the input volume with a user-defined vertical spacing $\mathcal{S}_\text{VG}$, e.g., of 10\,hPa.

\subsubsection{Computation of Vertical Front Line}
\label{sec:methods:frontalenvanalysis:verticalfrontline}

For the generation of our proposed novel types of vertical composites, we also need to compute the vertical front lines at $\mathcal{F}_{\text{HL}}(i)$.
Let $\mathcal{P}_{\text{V}}(i)$ denote the vertical plane that passes through $\mathcal{F}_{\text{HL}}(i)$ with $\mathcal{N}_{\text{HL}}(i)$ being parallel to the plane.
Then, similar to the horizontal front line, we can compute the vertical front line as the intersection $\mathcal{F}_{\text{VL}}(i) := \mathcal{F}_\text{S} \cap \mathcal{P}_{\text{V}}(i)$, represented as a polyline and sampled with a regular spacing $\mathcal{S}_\text{VL}$.
At a certain pressure level $\mathcal{L_\text{p}}$, the position of the vertical front line is then given on the vertical slice by 
$\mathcal{F}_{\text{VL}}(i,\mathcal{L}_{\text{p}})$.
In order to allow for an easy processing of the vertical front lines, we parametrize them in the 2D plane with coordinates $(u,v)$, centered horizontally at $\mathcal{F}_\text{VL}(i,\mathcal{L}_{\text{p}_{0}})$.
The parametrized vertical front line can then be defined as $\mathcal{F}^\text{uv}_{\text{VL}}(i,\mathcal{L}_{\text{p}}) =(u_i(\mathcal{L}_{\text{p}}), v(\mathcal{L}_{\text{p}})) = (u_i(\mathcal{L}_{\text{p}}), \mathcal{L}_{\text{p}})$, 
where $u_i(\mathcal{L}_{\text{p}_{0}}) = 0$ and $u_i(\mathcal{L}_{\text{p}})$ satisfies 
$\mathcal{F}^\text{uv}_{\text{VL}}(i,\mathcal{L}_{\text{p}}) = \mathcal{F}^\text{uv}_{\text{VL}}(i,\mathcal{L}_{\text{p}_{0}}) + (\mathcal{L}_{\text{p}}-\mathcal{L}_{\text{p}_{0}}) \mathbf{e}_z + u_i(\mathcal{L}_{\text{p}}) \mathbf{u}_i$.

\subsubsection{Computation of Vertical Front Line Means}
\label{sec:methods:frontalenvanalysis:verticalfrontlinemeans}

Let $\mathbb{F}^{\text{uv}}_{\text{VL}}=\{\mathcal{F}^{\text{uv}}_{\text{VL}}(i) \mid i\in \{0, \dots , n\}\},\,n\in \mathbb{N}$, be the set of $(u,v)$-parametrized vertical front lines of an individual front surface, computed for each point $\mathcal{F}_{\text{HL}}(i)$ of the horizontal front line $\mathcal{F}_{\text{HL}}$ at reference pressure level $\mathcal{L}_{\text{p}_0}$ and let furthermore $\mathcal{L}_{\text{p}_j} \in \{\mathcal{L}_{\text{p}_{min}}, \dots , \mathcal{L}_{\text{p}_{-1}},\mathcal{L}_{\text{p}_{0}},\mathcal{L}_{\text{p}_{1}},\dots,\mathcal{L}_{\text{p}_{max}}\}$ be a pressure level within the vertical range covered by the front surface.
Then, we compute the $(u,v)$-parametrized mean vertical front position
\begin{align}
\bar{\mathcal{F}}^{\text{uv}}_{\text{VL}}(\mathcal{L}_{\text{p}_j}) 
&= \frac{1}{N_j} \sum_{i \in I_j}{\mathcal{F}^{\text{uv}}_{\text{VL}}(i,\mathcal{L}_{\text{p}_j})}\\
&= \frac{1}{N_j} \sum_{i \in I_j}{\mathcal{F}^\text{uv}_{\text{VL}}(i,\mathcal{L}_{\text{p}_{0}}) + (\mathcal{L}_{\text{p}_j}-\mathcal{L}_{\text{p}_{0}}) \mathbf{e}_z + u_i(\mathcal{L}_{\text{p}_j}) \mathbf{u}_i}\\
&= \frac{1}{N_j} \sum_{i \in I_j}{\mathcal{L}_{\text{p}_{j}} \mathbf{e}_z + u_i(\mathcal{L}_{\text{p}_j}) \mathbf{u}_i},
\end{align}
where 
\[ I_j = \{i \in \{0, \dots, n\} \mid \mathcal{F}_\text{VL}(i,\mathcal{L}_{\text{p}_j})\, \text{exists}\}\]
is the set of indices for which vertical front lines are defined at the respective pressure level, and $N_j=|I_j|$ the number of these indices.

Then, for a set of front surfaces $\mathbb{F}_{\text{S}}=\{\mathcal{F}_{\text{S}}^{(k)} \mid k\in \{0, \dots , m\}\},\,m\in \mathbb{N}$ of which each member has a corresponding mean vertical front line $\mathcal{F}_{\text{VL}}^{(k)}$ with parametrization $\bar{\mathcal{F}}^{(k),\text{uv}}_{\text{VL}}(i,\mathcal{L}_{\text{p}})$, the mean vertical front line position over all front surfaces in $(u,v)$~--~coordinates is computed by
\begin{align}
\bar{\bar{\mathcal{F}}}^{\text{uv}}_{\text{VL}}(\mathcal{L}_{\text{p}_j})
&= \frac{1}{M_j} \sum_{k \in K_j}{\bar{\mathcal{F}}^{(k),\text{uv}}_{\text{VL}}(\mathcal{L}_{\text{p}_j})} \\
&= \frac{1}{M_j} \sum_{k \in K_j}{\mathcal{L}_{\text{p}_{j}} \mathbf{e}_v + u_k(\mathcal{L}_{\text{p}_j}) \mathbf{e}_u},
\end{align}
where
\[
K_j = \{k \in \{0, \dots, m\} \mid \mathcal{F}_{\text{VL}}^{(k)}(\mathcal{L}_{\text{p}_j})\, \text{exists}\}
\]
is the set of front surface indices for which a mean vertical front line position is defined at the respective pressure level, $M_j=|K_j|$ the number of these indices, and $\mathbf{e}_u = (1,0)$, $\mathbf{e}_v = (0,1)$ denote the unit vectors along the $u$-- and $v$--axes, respectively.

\subsubsection{Computing the Mean of the Vertical Slices}
\label{sec:methods:frontalenvanalysis:mean}

\begin{figure}[t]
    \centering
    \includegraphics[width=1\linewidth, trim=0.3cm 1.4cm 10.6cm 0.6cm, clip]{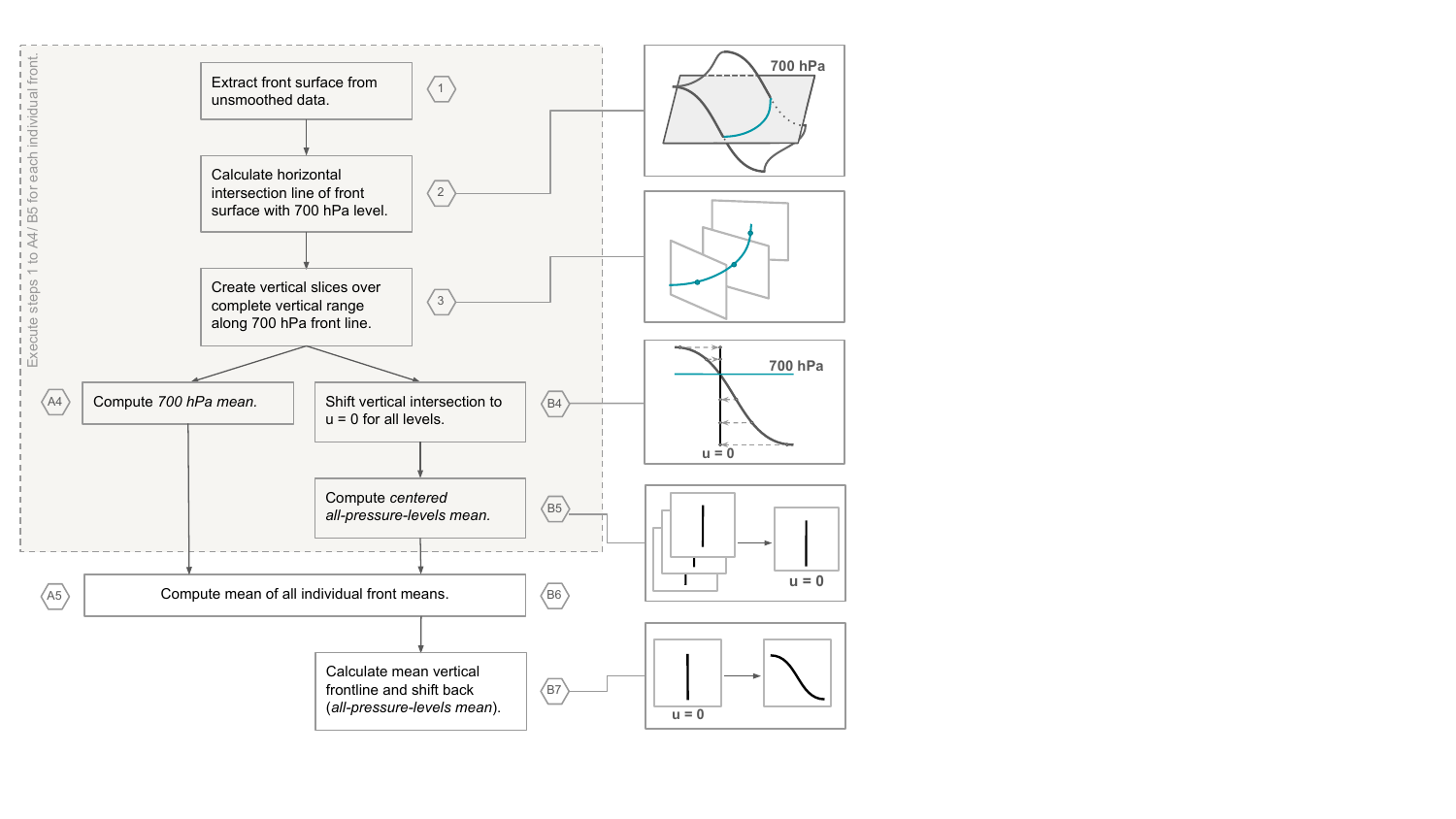}
    \caption{Workflow of the vertical composite computation using $700\,\text{hPa}$ as reference level. Steps 1~--~A5 illustrate the construction of the \emph{single-pressure-level mean}, while steps 1~--~B6 show the construction process of the \emph{centered all-pressure-levels mean}, from which the \emph{all-pressure-levels mean} is derived in the final step B7.}
    \label{fig:workflow}
\end{figure}

In order to obtain representative vertical cross-sections of the environment relative to the front, we compute the mean over all extracted vertical slices in two different ways.
The first way aligns the vertical slices w.r.t.\ to the front position in a single pressure level (\emph{single-pressure-level mean}) and is similar to the previous approach by Pacey et al.~\cite{paceyClimatologyNatureWarmseason2023}.
We present this approach here for comparison.
The second method, which represents our novel approach, aligns the fronts at all pressure levels (\emph{all-pressure-levels mean}).
Additionally, for the second method, we suggest the generation of two different visualizations, supporting different aspects of the front environment analysis, that is, either the analysis along the vertical (\emph{all-pressure-levels mean}) or the horizontal axis (\emph{centered all-pressure-levels mean}).
\begin{enumerate}
    \item [I.] The coordinate system of the vertical slice is defined such that the horizontal front line at $\mathcal{L}_{\text{p}_0}$ lies at $u=0\,\text{km}$.
          The mean is then computed over these vertical slices.
          Since the front positions will only be aligned at $\mathcal{L}_{\text{p}_0}$ and the slope of the vertical front line usually varies slightly along the horizontal front line, we expect that in this case, for all levels other than $\mathcal{L}_{\text{p}_0}$, we will average the values at positions with different distances to the front.
          We hypothesize that this way of computing the mean, which is similar to the method used to generate frontal cross sections in previous studies \cite{paceyClimatologyNatureWarmseason2023, schafferDriversColdFrontalPrecipitation2024}, will result in a blurred representation for most pressure levels, especially those further away from $\mathcal{L}_{\text{p}_0}$.
          We will refer to this representative mean as the \emph{single-pressure-level mean}.
    \item [II.a.] We shift the vertical slice values separately for each pressure level such that the entire front surface lies at $u=0\,\text{km}$ across all vertical levels. 
          The mean is then computed separately for each pressure level. 
          This way, the front surface is aligned across all slices and over all vertical levels.
          For pressure levels where no intersection with the front surface exists, values are treated as missing at that level.
          We count the number of slices that contribute to the mean of a specific vertical level in order to be able to correct for this imbalance.
          We will refer to this representative mean as the \emph{centered} all-pressure-levels mean.
    \item [II.b.] Starting from the \emph{centered all-pressure-levels mean}, for each pressure level, we shift the centered vertical slice values back in such a way that the value of the \emph{centered all-pressure-levels mean} at 0\,km matches the position of the mean vertical front line.
          This representation allows for a better comparison with the \emph{single-pressure-level mean} since the front structure will have a more natural appearance.
          It also allows for a better visual analysis of vertical relations of the atmospheric variables.
          We will refer to this representative mean as the \emph{all-pressure-levels mean}.
\end{enumerate}

We first compute the mean cross-sections for each front case separately.
Note that all of the computations described in the following are done in the $\text{uv}$--parametrization space, that is, similarly to Section~\ref{sec:methods:frontalenvanalysis:verticalfrontlinemeans}, we assume a $\text{uv}$--parametrization such that the front at $\mathcal{L}_{\text{p}_{0}}$ is located at $u=0\,\text{km}$.
If one is interested in a statistical analysis of multiple fronts, for I and II.a, we can simply compute the mean of the individual front-mean cross-sections.
For computing the representation of II.b, we first need to compute the mean of II.a and use this to compute the representation of II.b.
The workflow for constructing these composites is illustrated in Figure~\ref{fig:workflow}, in which steps 1~--~A5 lead to the representation of I, and steps 1~--~B6/B7 to the representations of II.a and II.b, respectively.

For the computation of the composites of multiple fronts, we decided to weight each front case equally.
An equal weighting seemed appropriate to us, as it does not favor long fronts, where usually the variation of the frontal environments along a single front varies only slightly.
However, alternatively the influence of the fronts could also be weighted according to their lengths.

An important aspect to take into account in the compositing approach based on 2D front surface is the vertically varying data availability.
For an individual front, the number of available data points varies between pressure levels, since the gradient of the equivalent potential temperature may be weaker in some regions, and hence the front surface will disappear in those regions.
We track individual data availability for each front by counting the number of data points at each level.
Furthermore, different fronts can cover different ranges of pressure levels, because the extracted front surfaces do not necessarily extend over the same vertical range.
Both aspects are taken into account in the average using only the available data at each pressure level and dividing by the corresponding number of data points.
In order to derive a representative mean frontal environment over all front cases, rather than a mean over all available data points, all fronts are equally weighted in this paper. 
Therefore, the data availability shown alongside the multiple-front means indicates the fraction of fronts contributing to the mean at the respective pressure level.

\begin{figure}[t]
    \centering

    \begin{subfigure}[b]{\linewidth}
        \centering

        \fbox{\raisebox{1.2cm}{\includegraphics[height=4.5cm, trim=3.81cm 1.76cm 3.65cm 2.65cm, clip]{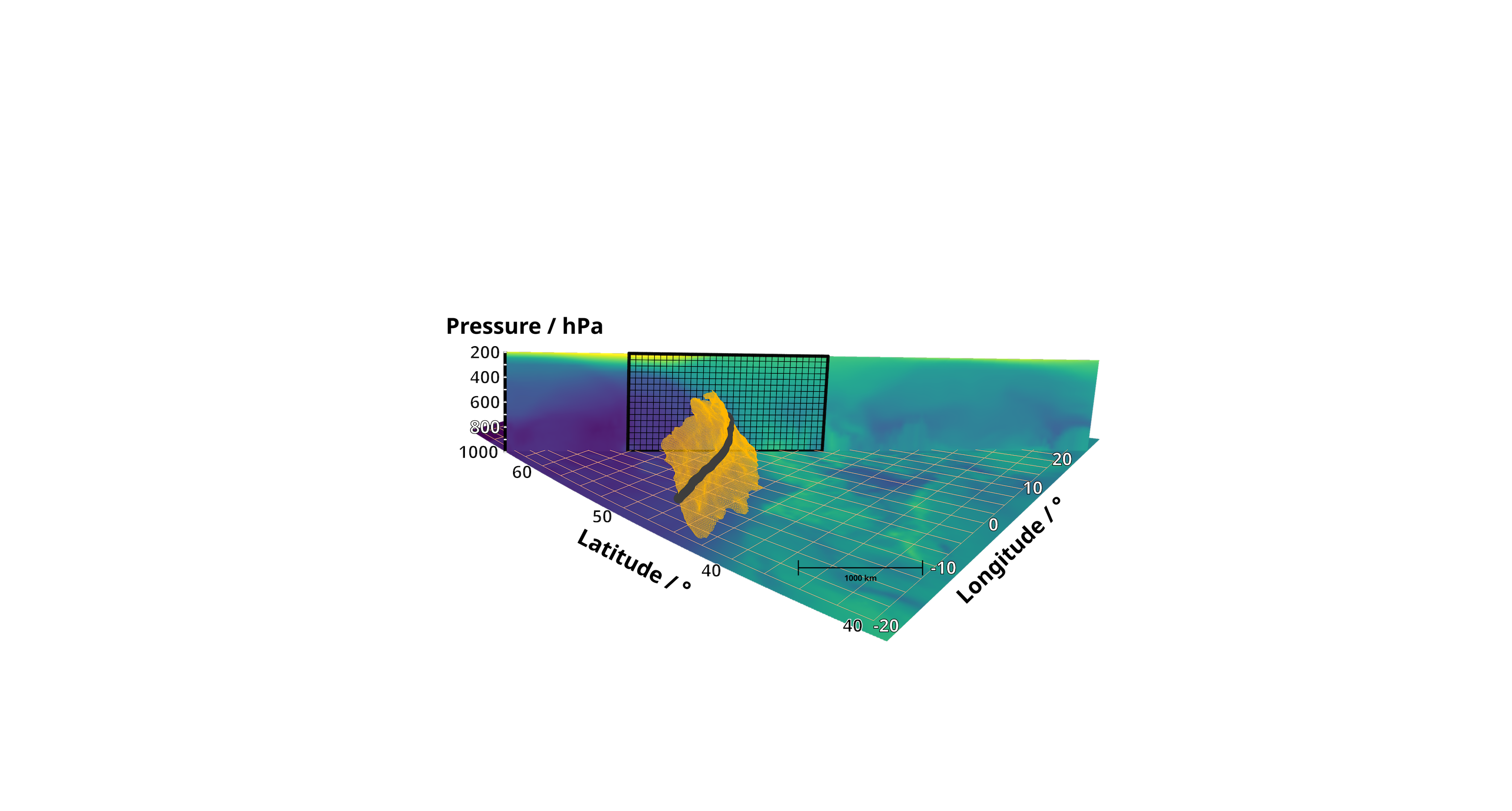}}
        \includegraphics[height=6.8cm, trim=0.23cm 0cm 0.13cm 0cm, clip]{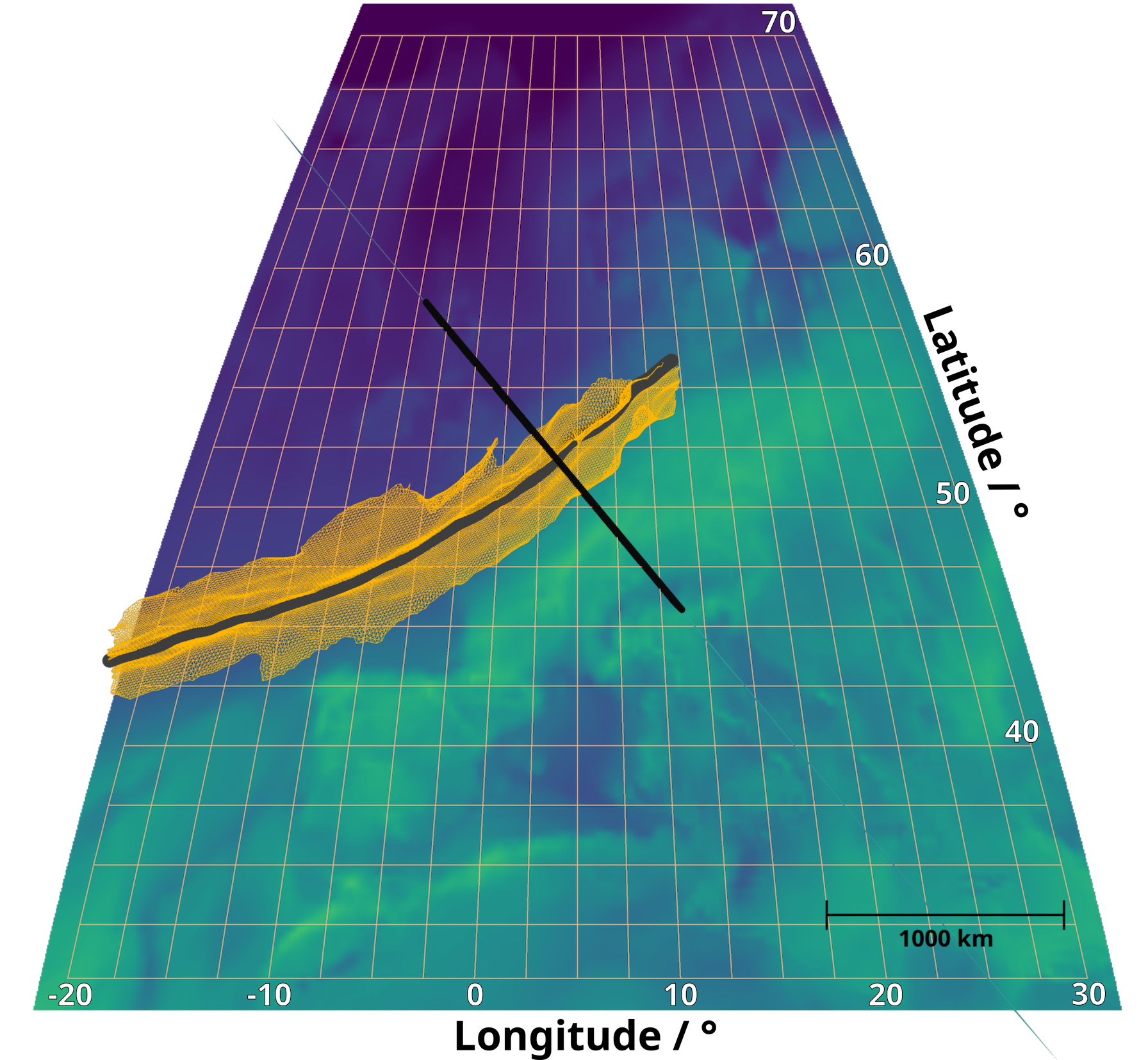} }
    \end{subfigure}

    \vspace{0.3cm}

    \begin{subfigure}[b]{0.5\linewidth}
        \centering
    \includegraphics[width=\linewidth]{Colormap_eqpt_280-360.png}

        \end{subfigure}
    \caption{Single cold front case of 11 May 2012. Side view (left) and top view (right) of the extracted 2D front surface (yellow triangular mesh), with sampled horizontal front line $\mathcal{F}_{\text{HL}}$ at 700\,hPa (dark gray), and grid G of the vertical slice (black, grid spacing of 50\,km and 50\,hPa for illustration) on which the field is evaluated at one point of the front line. The 3D scalar field of $\theta_e$ is represented by a horizontal cross-section at 1000\,hPa and a vertical cross-section.}
    \label{fig:spine_and_grid}
\end{figure}

\section{Results}
\label{sec:results}

In this section, the proposed compositing approach is first illustrated for a single cold front (Sect.~\ref{sec:results:casestudy}).
Subsequently, we present some preliminary results of our study on 50 cold fronts (Sect.~\ref{sec:results:multiplefronts}).
We only used synoptic-scale fronts of at least 1000\,km length.
The analysis is primarily intended to demonstrate the advantages of the proposed method rather than to provide a comprehensive climatological assessment. 
Throughout both parts, vertical cross-frontal composites of thermodynamic and kinematic fields based on the front surface (\emph{all-pressure-levels mean}) as determined by the ridge surface extraction method are compared to composites constructed relative to the one-dimensional frontline at a single pressure level (\emph{single-pressure-level mean}).
In this work, we used $\mathcal{L}_{\text{p}_0} = 700\,\text{hPa}$ to calculate the means.
For simplicity, in the following, we will refer to the \emph{single-pressure-level mean} as \emph{700\,hPa mean}.
In addition, we show composites in which the front position is centered at each pressure level, referred to as \emph{centered all-pressure-levels mean}, which provide complementary information on structures in the frontal environment by taking their distance to the front into account.

For all composites shown in this section, the parameters (Sect.~\ref{sec:methods:frontalenvanalysis}) are set to $\mathcal{S}_\text{HL}=25\,\text{km}$, $\mathcal{E}_\text{HG}=\pm 550\,\text{km}$, $\mathcal{S}_\text{HG}=10\,\text{km}$, and $\mathcal{S}_\text{VG}=\mathcal{S}_\text{VL}=10\,\text{hPa}$.
For composites based on the front surface, pressure levels are only considered in the calculation of the individual front means if at least 10\,\% of the number of data points at the reference level $\mathcal{L}_{\text{p}_0} = 700\,\text{hPa}$ are available.
For the multiple-front means, pressure levels are only shown if at least 10\,\% of the fronts, that is, at least five fronts, contribute to the mean.
The corresponding data availability is displayed alongside the composites, with 100\,\% corresponding to the data availability at the reference level $\mathcal{L}_{\text{p}_0} = 700\,\text{hPa}$.

\subsection{Single cold front example}
\label{sec:results:casestudy}

In this introductory example, our methods described in Section~\ref{sec:methods} are applied to the single cold front case of 11 May 2012 shown in Figure~\ref{fig:spine_and_grid}.
First, vertical cross-frontal composites are constructed for the 3D field used to extract the front surface, the magnitude of the horizontal gradient of $\theta_e$, to show how the frontal structure, characterized by strong horizontal gradients in the equivalent potential temperature, is represented in the respective composites.
The resulting composites are shown in Figure~\ref{fig:case_study_gradient}, with the \emph{700\,hPa mean} in Figure~\ref{fig:case_study_gradient_a} and the \emph{all-pressure-levels mean} in Figure~\ref{fig:case_study_gradient_b}.
For the \emph{700\,hPa mean}, the analysed domain extends from 1000 to 200\,hPa.
For the \emph{all-pressure-levels mean}, only pressure levels at which a front surface was extracted are included, resulting in a vertical range of 970\,--\,470\,hPa.
The bar next to the \emph{all-pressure-levels mean} indicates the number of data points available at each pressure level relative to the maximum number at 700\,hPa.
In both composites, the mean vertical front line is shown in black as a reference.
We can observe that in both composites the strongest gradients are centered close to the mean vertical front line and form a narrow, nearly symmetric band around it.
Maximum values occur at around 800\,hPa in both composites, but they are considerably higher in the \emph{all-pressure-levels mean}.
Furthermore, in the \emph{all-pressure-levels mean}, the maximum at each pressure level is located exactly on the mean vertical front line, because each level is composited relative to the corresponding extracted front position.
In the \emph{700\,hPa mean}, this is only guaranteed at 700\,hPa.
At levels below 700\,hPa, the maximum is partly shifted away from the reference line.
Above 700\,hPa, a second region of enhanced gradient is visible behind the mean vertical front line. 
This feature presumably represents an upper-level cold front that does not extend to the Earth's surface. 
By extracting the front as a 2D manifold, we can distinguish the front that extends from the surface into the upper levels from such upper-level fronts.
If the front had been extracted as a front line from smoothed data at a pressure level above 700\,hPa, the resulting line might have been located further towards the second frontal region and would therefore not have represented the exact position of the front.
This shows that it is important to consider the front as a surface for the calculation of composites rather than just using a single front line at a particular pressure level.
It also ensures that distances in the composite are measured relative to the actual front position at each pressure level.

\begin{figure}[hb]
    \centering
    \begin{subfigure}[b]{0.48\linewidth}
        \centering
        \includegraphics[height=6cm]{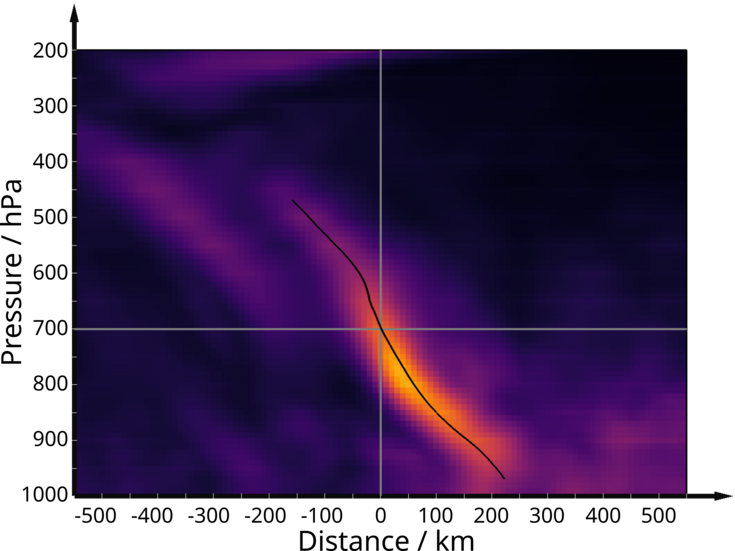} 
        \caption{} 
        \label{fig:case_study_gradient_a}
    \end{subfigure}%
    \hfill%
    \begin{subfigure}[b]{0.52\linewidth}
        \begin{minipage}[c][6cm][c]{0.89\linewidth}
        \centering
        \includegraphics[height=6cm, trim=0.35cm 0cm 0cm 0cm, clip]{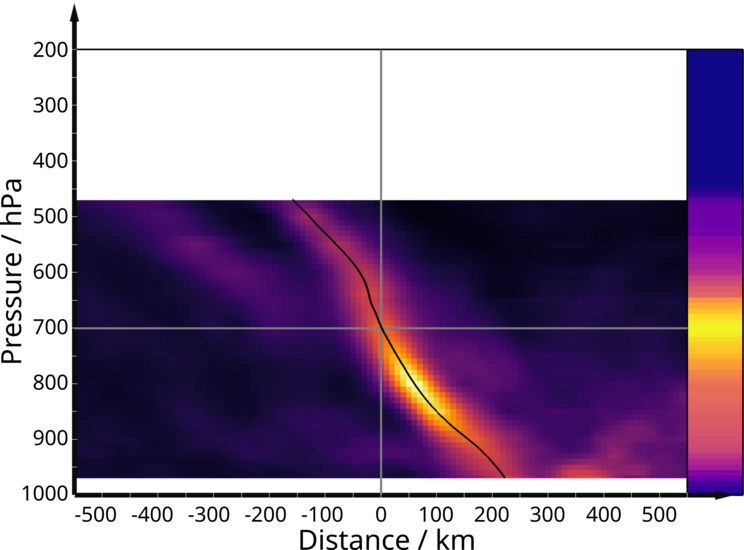}    
        \end{minipage}%
        \begin{minipage}[c][6cm][c]{0.11\linewidth}
            \centering
            \includegraphics[angle=90, height=5 cm]{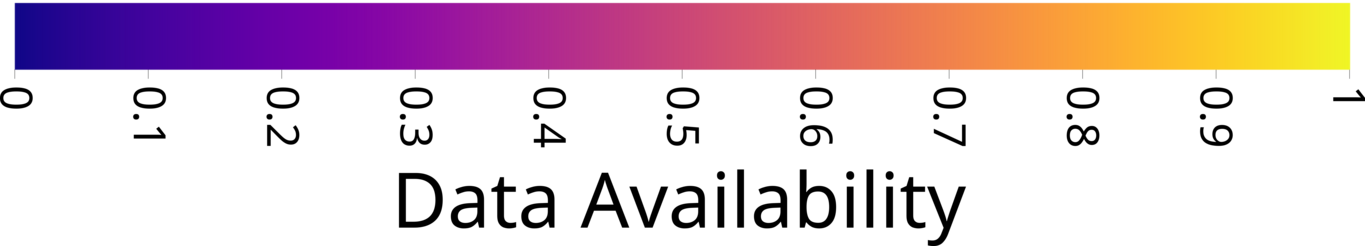}
        \end{minipage}
        
        \caption{} 
        \label{fig:case_study_gradient_b}
    \end{subfigure}

    \vspace{0.3cm}

    \begin{subfigure}[b]{0.45\linewidth}
        \hfill
    \includegraphics[width=\linewidth, trim=0cm 0cm 0cm 0cm, clip]{Colormap_gradient_0-21.png}

    \end{subfigure}
 
    \caption{Single-front mean of the horizontal gradient magnitude of $\theta_e$ averaged relative to the frontal position. (a)~\emph{700\,hPa mean}, that is, with the frontal position aligned to \(u=0\,km\) at 700\,hPa; (b)~\emph{all-pressure-levels mean}, that is, with the frontal position aligned to mean frontal position (displayed by the black line) at all levels. The fraction of available data for the \emph{all-pressure-levels mean} is indicated on the right.}
    \label{fig:case_study_gradient}
\end{figure}

The corresponding \emph{700\,hPa mean} and \emph{all-pressure-levels mean} composites for the equivalent potential temperature, vertical velocity, and convergence fields are shown in Figure~\ref{fig:case_study_var_fields}.
These fields are used to show how the choice of reference geometry affects the representation of thermodynamic and kinematic structures in the frontal environment.
While the equivalent potential temperature characterizes the thermodynamic properties of the air masses on both sides of the front, vertical velocity and convergence describe the atmospheric flow in its vicinity.

\begin{figure}[H]
    \centering

    \begin{minipage}[c]{0.94\linewidth}
        \centering
        \begin{subfigure}[b]{0.495\linewidth}
            \centering
            \includegraphics[height=5.9cm, trim=0cm 0cm 0cm 0cm, clip]{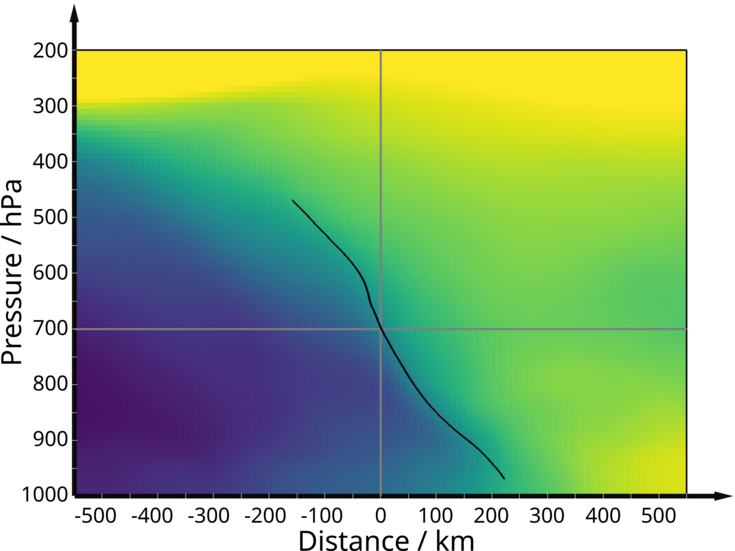}
            \caption{} 
            \label{fig:case_study_var_fields_a}
        \end{subfigure}
        \hfill%
        \begin{subfigure}[b]{0.495\linewidth}
            \centering
            \includegraphics[height=5.9cm, trim=0.35cm 0cm 0cm 0cm, clip]{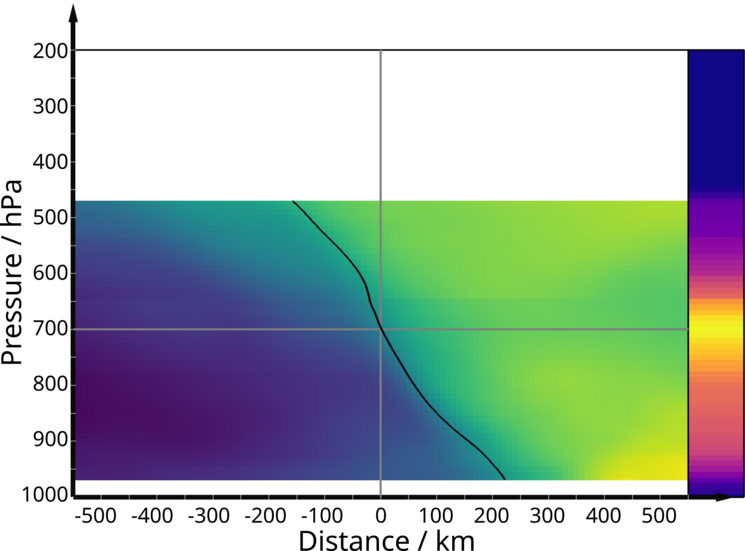} 
            \caption{} 
            \label{fig:case_study_var_fields_b}
        \end{subfigure}
        
        \begin{subfigure}[b]{0.45\linewidth}
            \centering
        \includegraphics[width=\linewidth]{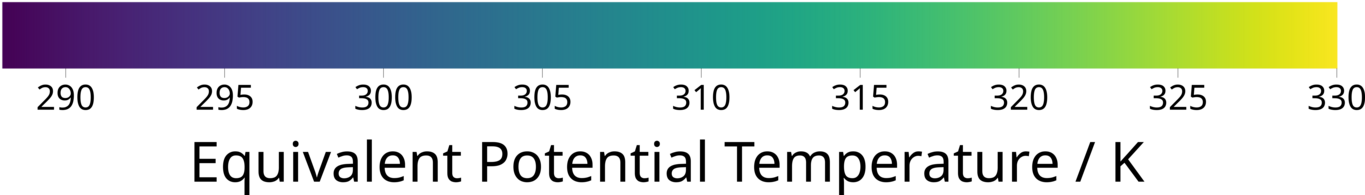}
        \end{subfigure}%
        
        \vspace{0.1cm}
        \begin{subfigure}[b]{0.495\linewidth}
            \centering
            \includegraphics[height=5.9cm, trim=0cm 0cm 0cm 0cm, clip]{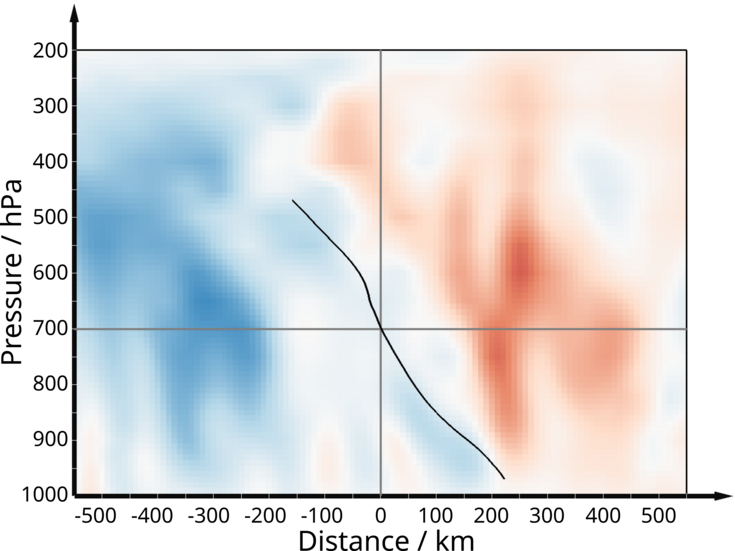} 
            \caption{} 
            \label{fig:case_study_var_fields_c}
        \end{subfigure}
        \hfill%
        \begin{subfigure}[b]{0.495\linewidth}
            \centering
            \includegraphics[height=5.9cm, trim=0.35cm 0cm 0cm 0cm, clip]{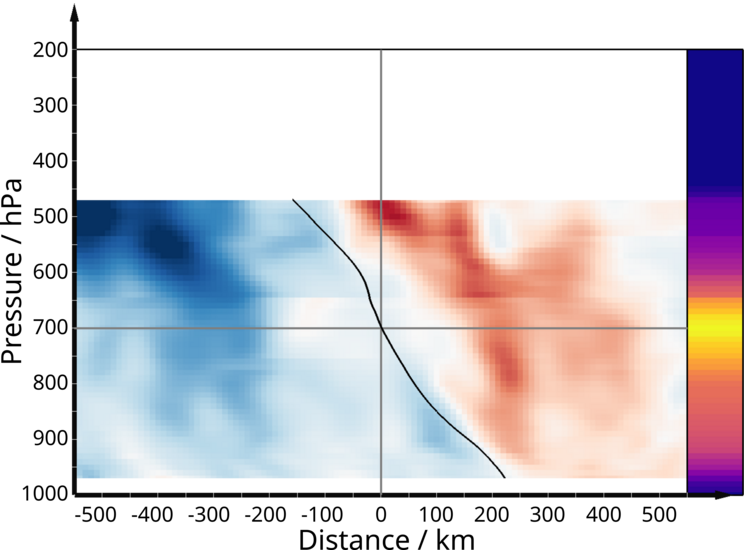} 
            \caption{} 
            \label{fig:case_study_var_fields_d}
        \end{subfigure}
        \begin{subfigure}[b]{0.45\linewidth}
            \centering
        \includegraphics[width=\linewidth]{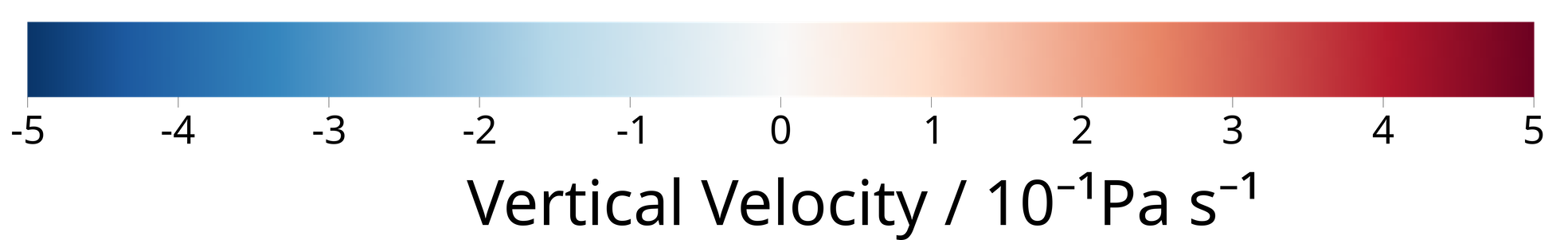}   
        \end{subfigure}  
        
        \vspace{0.1cm}
        \begin{subfigure}[b]{0.495\linewidth}
            \centering
            \includegraphics[height=5.9cm, trim=0cm 0cm 0cm 0cm, clip]{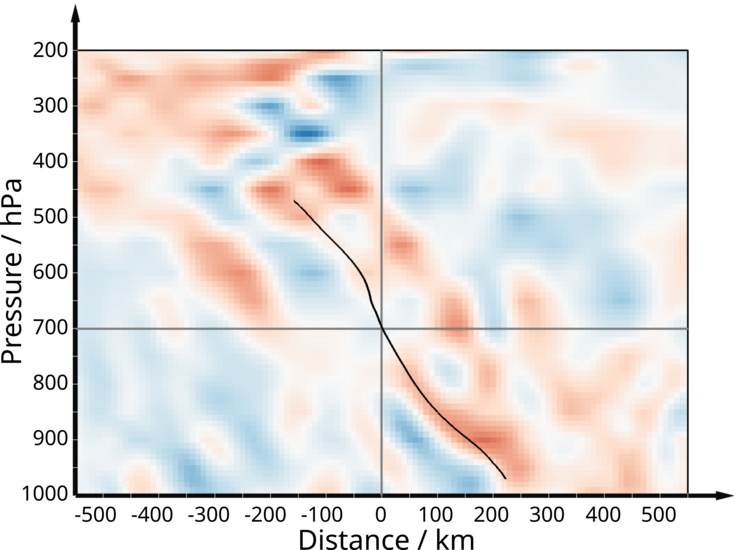} 
            \caption{} 
            \label{fig:case_study_var_fields_e}
        \end{subfigure}
        \hfill
        \begin{subfigure}[b]{0.495\linewidth}
            \centering
            \includegraphics[height=5.9cm, trim=0.35cm 0cm 0cm 0cm, clip]{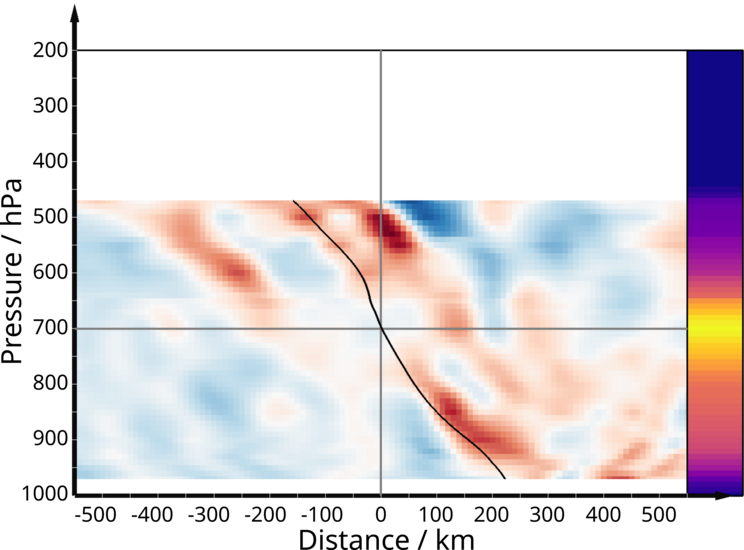} 
            \caption{} 
            \label{fig:case_study_var_fields_f}
        \end{subfigure}
        \begin{subfigure}[b]{0.45\linewidth}
            \centering
        \includegraphics[width=\linewidth]{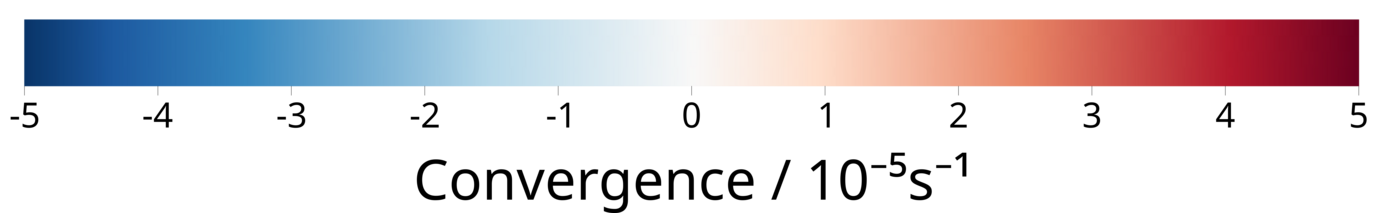}
        \end{subfigure}
    \end{minipage}%
    \hfill%
    \begin{minipage}[c]{0.059\linewidth}
        \centering
        \includegraphics[angle=90]{Colormap_dataAvailability_vertical.png}
    \end{minipage}
\caption{Single-front mean equivalent potential temperature, vertical velocity, and convergence fields averaged relative to the frontal position. (a,c,e)~\emph{700\,hPa mean}, that is, with the frontal position aligned only at 700\,hPa; (b,d,f)~\emph{all-pressure-levels mean}, that is, with the frontal position aligned at all levels.
The black line indicates the mean frontal position. 
Positive values of the vertical velocity indicate upward movement. 
The fraction of available data for the \emph{all-pressure-levels mean} is indicated on the right.}

\label{fig:case_study_var_fields}
\end{figure}

Since the latter two variables are derived from the wind field and not from the thermodynamic field that is used for the front detection, these variables provide a complementary perspective on the frontal environment in our analyses.

In both composites of $\theta_e$, the colder air mass forms a wedge-shaped structure beneath the warmer air.
The mean front line clearly separates the cold and comparatively dry air behind the front from the warm and moist air ahead of it.
The strongest thermodynamic contrast is located along the front line in both composites below 700\,hPa.
Above 700\,hPa, the contrast is less pronounced, while a second wedge-shaped structure towards the colder air mass becomes apparent, as already seen in the corresponding composites of the horizontal gradient magnitude of $\theta_e$.
Although the differences between both compositing approaches are subtle for this particular front case, the contrast along the mean vertical frontline is slightly sharper in the \emph{all-pressure-levels mean} (Fig.~\ref{fig:case_study_var_fields_b}).

For the kinematic variables, the differences between the two compositing approaches are more pronounced.
In the vertical velocity field, both composites show ascending air ahead of the front and descending air behind it, as expected for a cold front~\cite{martinFronts2024}. 
However, the signal is stronger in the \emph{all-pressure-levels mean}, in particular at upper levels.
In this composite, additional structures become visible close to the front line on the warm-air side between 800 and 900\,hPa. 
Overall, upward and downward motion appear to be more clearly separated by the front line in the \emph{all-pressure-levels mean}, which suggests that the vertical circulation associated with the front is more coherently represented when each pressure level is aligned with the corresponding front position.

Figure~\ref{fig:case_study_var_fields_e} and \ref{fig:case_study_var_fields_f} show the composites for the horizontal convergence field.
Associated with ageostrophic cross-frontal circulation~\cite{martinFronts2024}, we expect lower-level convergence and upper-level divergence in the regions of ascent ahead of the front and the opposite signal in the descending cold air.
Such structures are observed in both \emph{700\,hPa mean} and \emph{all-pressure-levels mean}, albeit with some noise. 
In the \emph{all-pressure-levels mean} (Fig.~\ref{fig:case_study_var_fields_f}), positive values are greater than in the \emph{700\,hPa mean} (see corresponding histograms in Fig.~\ref{fig:histograms_vars} in Appendix~\ref{sec:compositehistograms}), especially ahead of the front close to the mean front line at lower levels as well as at upper levels.
Enhanced divergence is also visible in the \emph{all-pressure-levels mean}, located almost vertically above the region of enhanced lower-level convergence.
Overall, the vertical structure becomes clearer in the \emph{all-pressure-levels mean}.

\subsection{Preliminary Statistics Across Multiple Fronts}
\label{sec:results:multiplefronts}

\begin{figure}[htb]
    \centering

    \begin{subfigure}[b]{0.48\linewidth}
        \centering
        \includegraphics[height=6cm]{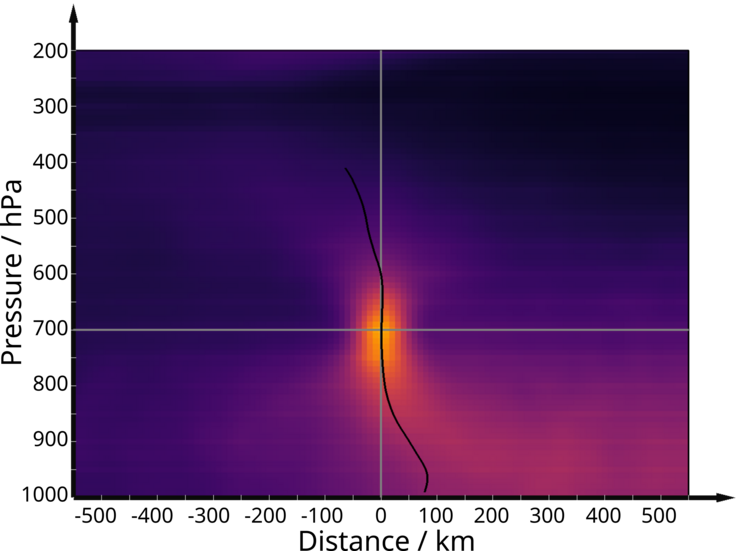} 
        \caption{} 
        \label{fig:multi_front_gradient_a}
    \end{subfigure}%
    \hfill%
    \begin{subfigure}[b]{0.52\linewidth}
        \begin{minipage}[c][6cm][c]{0.89\linewidth}
        \centering
        \includegraphics[height=6cm, trim=0.35cm 0cm 0cm 0cm, clip]{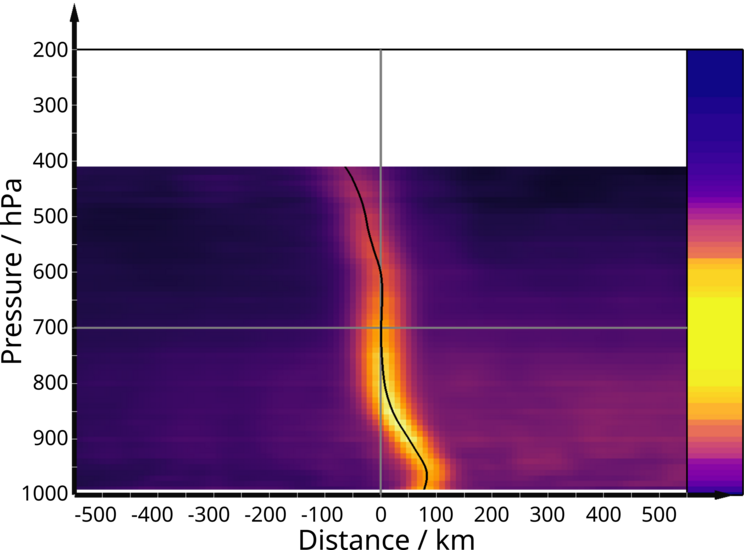}    
        \end{minipage}%
        \begin{minipage}[c][6cm][c]{0.11\linewidth}
            \centering
            \includegraphics[angle=90, height=5 cm]{Colormap_dataAvailability_vertical.png}
        \end{minipage}
        
        \caption{} 
        \label{fig:multi_front_gradient_b}
    \end{subfigure}

    \vspace{0.3cm}

    \begin{subfigure}[b]{0.45\linewidth}
        \hfill
    \includegraphics[width=\linewidth, trim=0cm 0cm 0cm 0cm, clip]{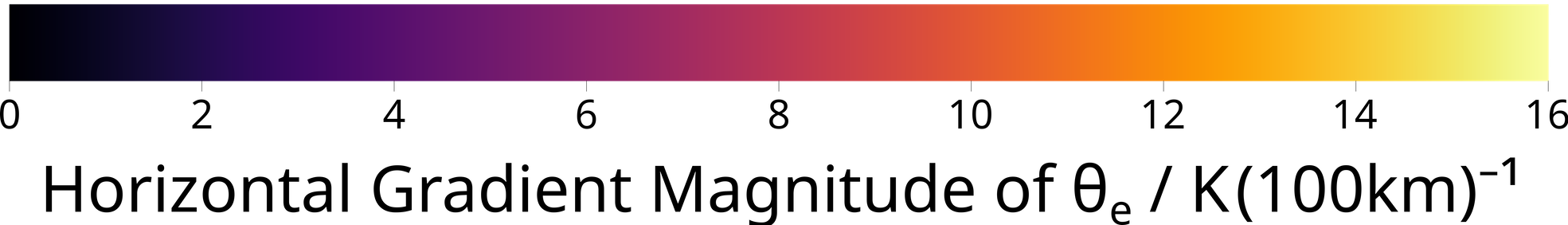}

    \end{subfigure}

    \caption{Multiple-front mean of the horizontal gradient magnitude of $\theta_e$ averaged relative to the frontal position. (a)~\emph{700\,hPa mean}, that is, with the frontal position aligned to \(u=0\,km\) at 700\,hPa; (b)~\emph{all-pressure-levels mean}, that is, with the frontal position aligned to mean frontal position (displayed by the black line) at all levels. The fraction of available data for the \emph{all-pressure-levels mean} is indicated on the right.}
    \label{fig:multi_front_gradient}
\end{figure}

In this section, we describe the results of applying the proposed compositing method described in Section~\ref{sec:methods} to a set of 50 cold fronts. A description of the data set and a complete list of the front cases are provided in Section~\ref{sec:data} and Table~\ref{tab:list_fronts}, respectively.

Similarly to the results for the single front presented in Section~\ref{sec:results:casestudy}, we computed the \emph{700\,hPa mean} and the \emph{centered all-pressure-levels mean} fields for each of the 50 fronts separately.
For each front, we ignored pressure levels in the \emph{centered all-pressure-levels mean} field if there was less than 10\,\% data available compared to the maximum number of possible cross-sectional data.
This resulted in the loss of very low and very high pressure levels.

In order to compute the \emph{700\,hPa means} and \emph{centered all-pressure-levels means} of multiple fronts, we only had to average the mean fields of all individual fronts as these were already normalized (centered) either w.r.t.\ the 700\,hPa front position only (\emph{700\,hPa mean}) or w.r.t.\ to the front position at all pressure levels (\emph{centered all-pressure-levels mean})(cf. Sect.~\ref{sec:methods:frontalenvanalysis:mean}).
However, similarly to the computation of the mean fields of the single fronts, we removed pressure levels if the number of single fronts contributing to this pressure level was less than 10\,\%, that is, for 50 fronts, at least five fronts had to contribute data to a pressure level to include this pressure level in the composite computation.
The resulting multiple-front \emph{centered all-pressure-levels mean} fields extend from 980 to 410\,hPa.
For most of the cold fronts studied, extracting the front surfaces was difficult both down to the surface level, that is, to the 1000\,hPa level, and above the 450\,hPa level.
Hence, the data density is comparatively low in the upper and lower pressure levels.
Finally, from the \emph{centered all-pressure-levels mean} fields, we then computed the \emph{all-pressure-levels mean} fields by shifting the individual pressure levels back to the mean front position, as described in Section~\ref{sec:methods:frontalenvanalysis:mean}.

Figure~\ref{fig:multi_front_gradient} shows the \emph{700\,hPa mean} and the \emph{all-pressure-levels mean} of the horizontal gradient magnitude of $\theta_e$.
For reference, the mean vertical front line of all 50 fronts, computed from the front surfaces, is displayed in both composites.
In comparison to the \emph{700\,hPa mean} of the gradient magnitude computed for the single front depicted in Figure~\ref{fig:case_study_gradient_a}, Figure~\ref{fig:multi_front_gradient_a} shows a much less clear structure.
The mean front line no longer fits to the underlying data except at the 700\,hPa pressure level.
The reduced clarity of the 700\,hPa composite is caused by differences in the slopes of the frontal surfaces. 
Although the horizontal front lines at 700\,hPa are aligned before averaging, the fronts extend upward and downward with different inclinations. 
Consequently, the composite averages values from locations that are at different distances to the actual front surface.
This blurs the gradient signal and produces a much less distinct structure.
In contrast to this, in the \emph{all-pressure-levels mean} (Fig.~\ref{fig:multi_front_gradient_b}), the strongest gradients are aligned with the mean front line at each pressure level.
This also supports the claim that the applied ridge surface extraction method works correctly and is capable of identifying the regions of largest $\theta_e$ gradient of the front at each pressure level.

A similar picture emerges for the equivalent potential temperature field. 
In the \emph{700\,hPa mean} of $\theta_e$ (Fig.~\ref{fig:multi_front_var_fields_a}), the separation between the two air masses is clearly visible only close to the 700\,hPa level, whereas the thermodynamic contrast becomes increasingly blurred above and below this reference level. 
In the \emph{all-pressure-levels mean} (Fig.~\ref{fig:multi_front_var_fields_b}), however, the cold and comparatively dry air behind the front and the warm and moist air ahead of the front are clearly separated along the mean front line over the given range of pressure levels. 
This indicates that the thermodynamic structure is better preserved when the front position is taken into account at each pressure level.

\begin{figure}[htbp]
    \centering

    \begin{minipage}[c]{0.94\linewidth}
        \centering

        \begin{subfigure}[b]{0.495\linewidth}
            \centering
            \includegraphics[height=5.9cm, trim=0cm 0cm 0cm 0cm, clip]{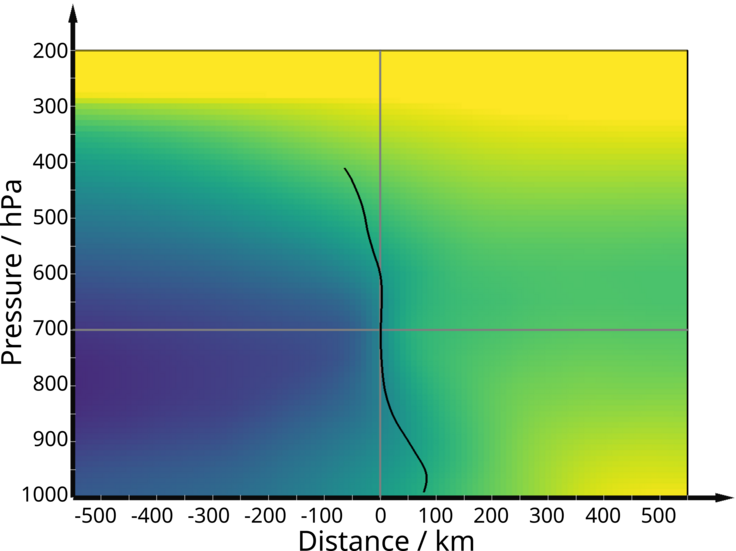}
            \caption{} 
            \label{fig:multi_front_var_fields_a}
        \end{subfigure}
        \hfill
        \begin{subfigure}[b]{0.495\linewidth}
            \centering
            \includegraphics[height=5.9cm, trim=0.35cm 0cm 0cm 0cm, clip]{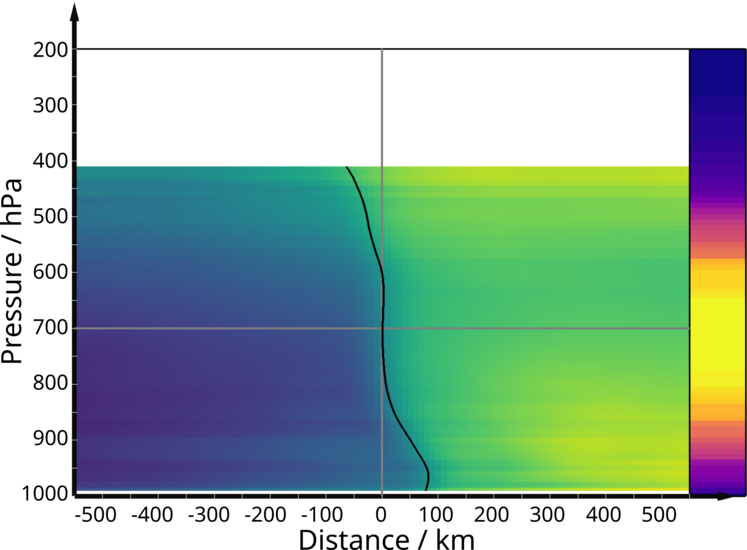}
            \caption{} 
            \label{fig:multi_front_var_fields_b}
        \end{subfigure}

        \begin{subfigure}[b]{0.45\linewidth}
            \centering
        \includegraphics[width=\linewidth]{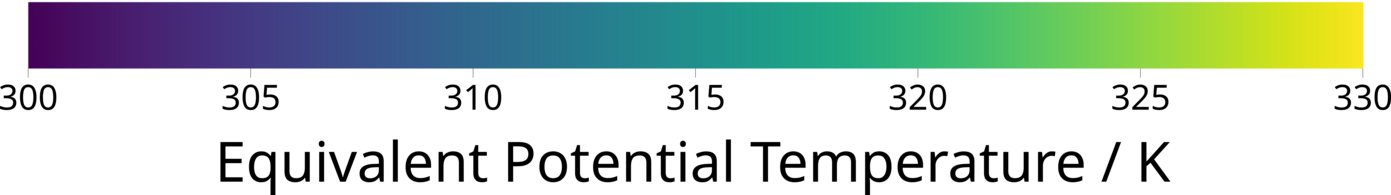}
    
        \end{subfigure}
        
        \vspace{0.1cm}
    
        \begin{subfigure}[b]{0.495\linewidth}
            \centering
            \includegraphics[height=5.9cm, trim=0cm 0cm 0cm 0cm, clip]{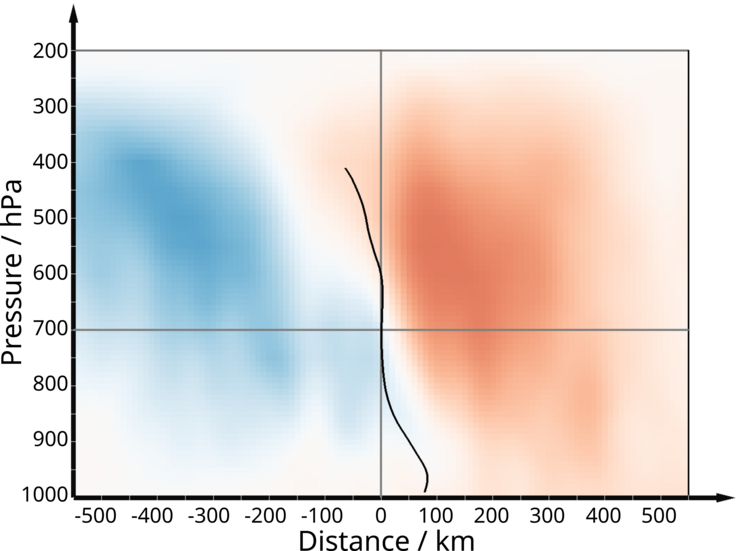} 
            \caption{} 
            \label{fig:multi_front_var_fields_c}
        \end{subfigure}
        \hfill
        \begin{subfigure}[b]{0.495\linewidth}
            \centering
            \includegraphics[height=5.9cm, trim=0.35cm 0cm 0cm 0cm, clip]{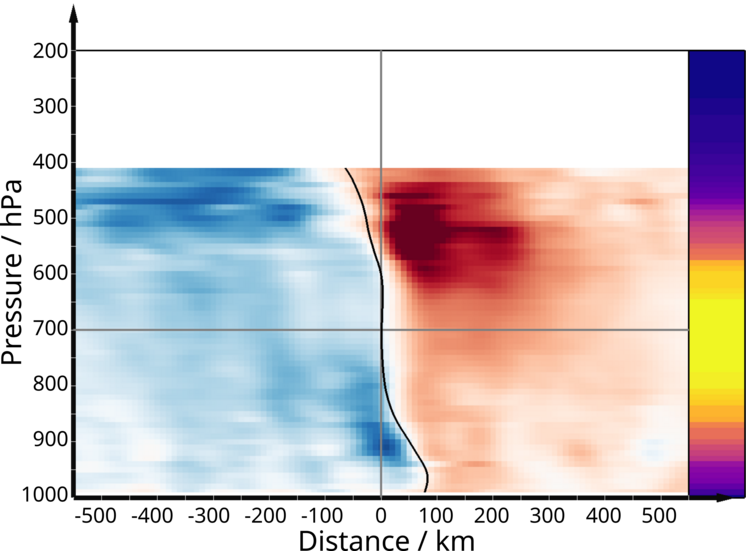} 
            \caption{} 
            \label{fig:multi_front_var_fields_d}
        \end{subfigure}

        \begin{subfigure}[b]{0.45\linewidth}
            \centering
        \includegraphics[width=\linewidth]{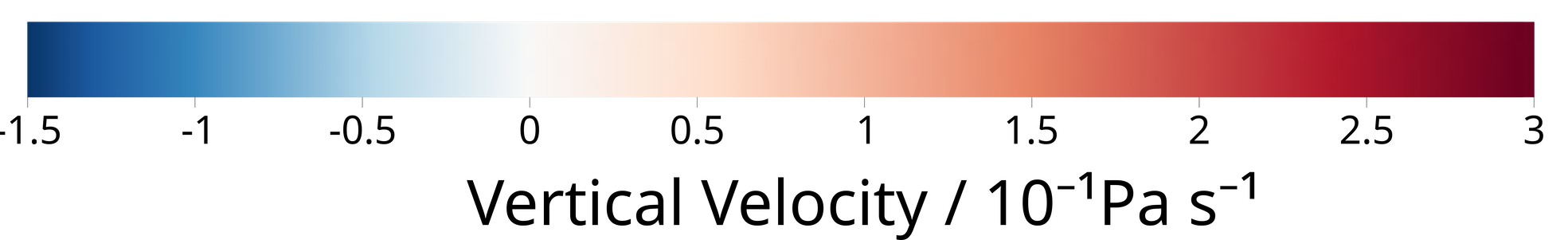}
    
        \end{subfigure}
        
        \vspace{0.1cm}

        \begin{subfigure}[b]{0.495\linewidth}
            \centering
            \includegraphics[height=5.9cm, trim=0cm 0cm 0cm 0cm, clip]{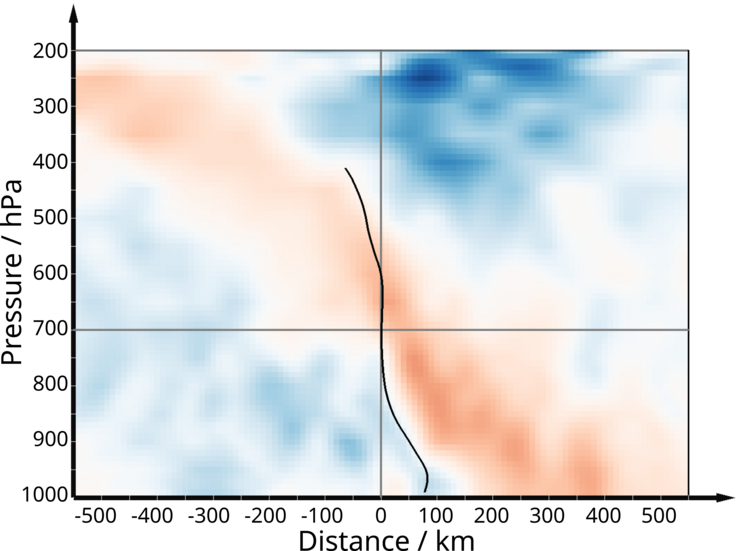} 
            \caption{} 
            \label{fig:multi_front_var_fields_e}
        \end{subfigure}
        \hfill
        \begin{subfigure}[b]{0.495\linewidth}
            \centering
            \includegraphics[height=5.9cm, trim=0.35cm 0cm 0cm 0cm, clip]{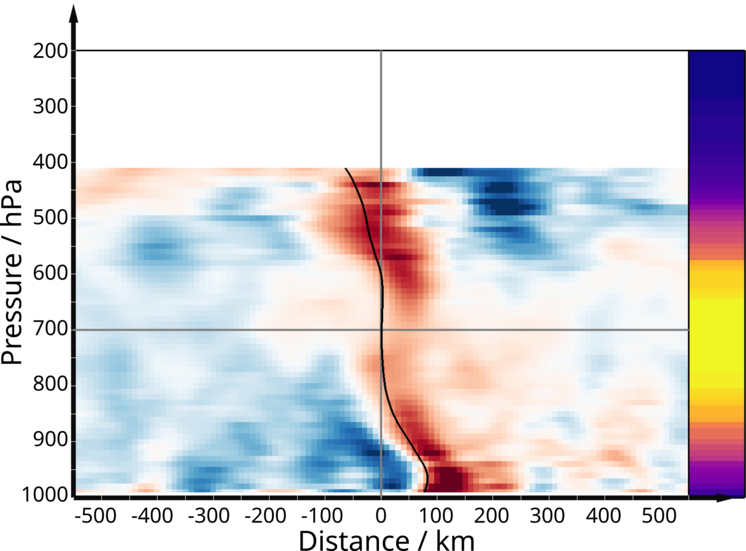} 
            \caption{} 
            \label{fig:multi_front_var_fields_f}
        \end{subfigure}

        \begin{subfigure}[b]{0.45\linewidth}
            \centering
        \includegraphics[width=\linewidth]{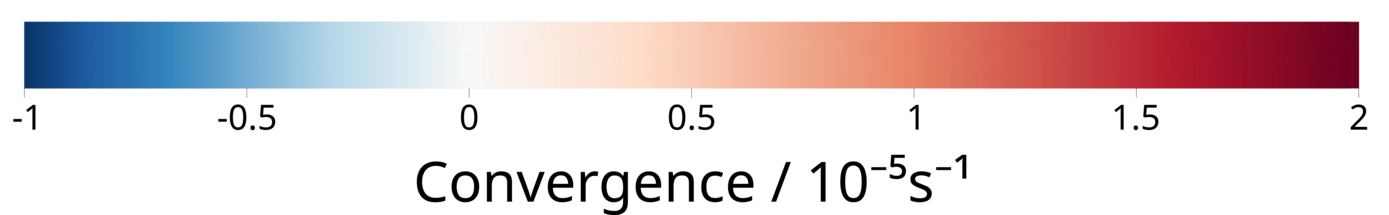}
    
        \end{subfigure}
        
    \end{minipage}%
    \hfill%
    \begin{minipage}[c]{0.06\linewidth}
        \centering
        \includegraphics[angle=90]{Colormap_dataAvailability_vertical.png}
    \end{minipage}

\caption{Multiple-front mean equivalent potential temperature, vertical velocity and convergence fields averaged relative to the frontal position. (a,c,e)~\emph{700\,hPa mean}, that is, with the frontal position aligned only at 700\,hPa; (b,d,f)~\emph{all-pressure-levels mean}, that is, with the frontal position aligned at all levels.
The black line indicates the mean frontal position. 
Positive values of the vertical velocity indicate upward movement. 
The fraction of available data for the \emph{all-pressure-levels mean} is indicated on the right.}

\label{fig:multi_front_var_fields}
\end{figure}

The differences between the two compositing approaches are similarly pronounced for the kinematic fields. 
The \emph{700\,hPa mean} of vertical velocity (Fig.~\ref{fig:multi_front_var_fields_c}) shows ascending air ahead of the front and descending air behind it, with the highest values mainly between about 750 and 400\,hPa. However, the signal is broad and no distinct maximum or minimum can be identified. 
Around the mean front line, values are close to zero over a large part of the vertical range, likely because the variability of the frontal slope between individual front cases causes regions of upward and downward motion to partially cancel each other when averaged relative to the 700\,hPa front position only.
In the \emph{all-pressure-levels mean} (Fig.~\ref{fig:multi_front_var_fields_d}), the vertical velocity field shows a much clearer separation between upward motion ahead of the front and downward motion behind it across all available pressure levels. 
A distinct maximum in upward motion is located directly ahead of the front between about 700 and 600\,hPa. 
Regions of greatest descent are also more clearly identifiable at higher levels above 500\,hPa, although the uncertainty increases there because less than half of the fronts contribute data at these levels.
In particular, a minimum in vertical velocity can be observed directly behind the front at around 900\,hPa, which is not visible in the \emph{700\,hPa mean}.
This might be the low-level signature of dry intrusions, dry air streams that typically descend behind cold fronts~\cite{carlson1980airflow}.%

The convergence field shows a consistent picture. 
In the \emph{all-pressure-levels mean} (Fig.~\ref{fig:multi_front_var_fields_f}), convergence is strongest directly ahead of the front at all levels, but in particular at lower levels, where it is consistent with the onset of upward motion. 
Divergence at surface level behind the front is associated with descending air. 
Compared to the \emph{700\,hPa mean} (Fig.~\ref{fig:multi_front_var_fields_e}), the convergence and divergence patterns are more clearly related to the front line.

Overall, the signals away from the 700\,hPa level are much stronger in the \emph{all-pressure-levels mean} (see also the corresponding histograms in Fig.~\ref{fig:histograms_vars} in the appendix), indicating that the kinematic structures are better preserved when the fronts are aligned level by level.
Furthermore, the kinematic fields in the \emph{all-pressure-levels mean} are more consistent with the thermodynamic structure in the frontal environment.

In the \emph{centered all-pressure-levels mean}, the front is located at $u=0\,km$ at all levels, making equidistant features in the frontal environment immediately visually identifiable.
The resulting composites of all fields considered are shown in Figure~\ref{fig:multi_front_centered}.
The presented data are the same as in the \emph{all-pressure-levels mean} but shifted according to their distance to the front.
Therefore, $\lVert \nabla_h \theta_e \rVert$ is maximal at $u=0\,km$ at each pressure level (see Fig.~\ref{fig:multi_front_centered_a}).
For $\theta_e$, the air masses are clearly separated at the frontal position throughout the vertical column.
In Figure~\ref{fig:multi_front_centered_c}, the maximum upward motion at each pressure level is located within the first 100~--~200\,km on the warm-air side of the front.
A similar pattern is found for convergence, with the strongest convergence at each pressure level also occurring within this distance ahead of the front.

\begin{figure}[h]
    \centering
    \begin{minipage}[c]{0.94\linewidth}

    \begin{subfigure}[b]{0.495\linewidth}
        \centering
            \includegraphics[height=6cm,, trim=0.0cm 0cm 0cm 0cm, clip]{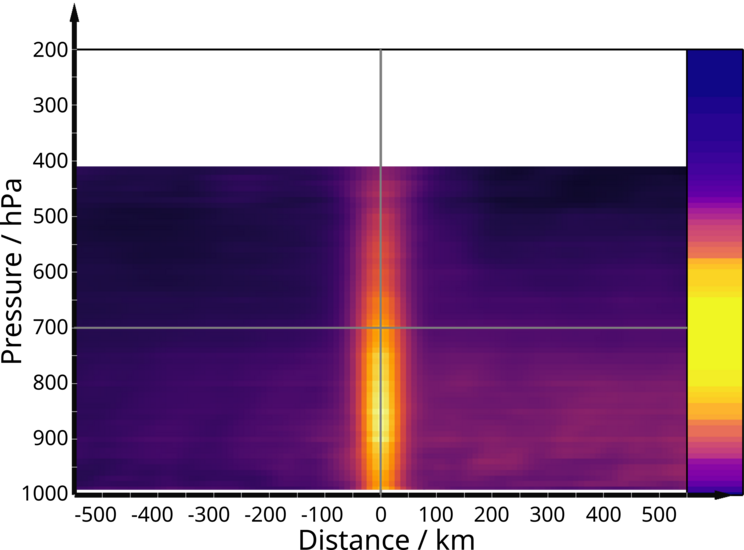}            
    \vspace{0.3cm}  
    \centering
        \hfill
        \includegraphics[width=0.9\linewidth]{Colormap_gradient_0-16.png}
        \caption{} 
        \label{fig:multi_front_centered_a}
    \end{subfigure}   
    \hfill
    \begin{subfigure}[b]{0.495\linewidth}
        \centering
        \includegraphics[height=6cm, trim=0.35cm 0cm 0cm 0cm, clip]{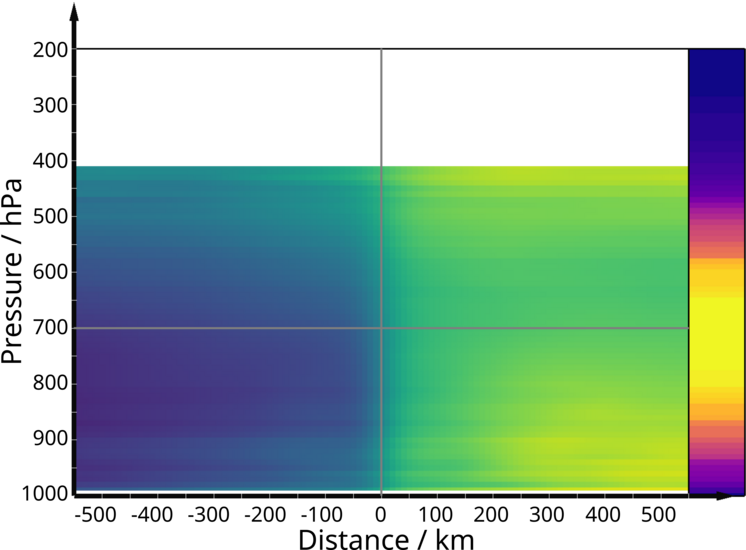}  
    \vspace{0.3cm}   
    \centering
    \includegraphics[width=0.9\linewidth]{Colormap_eqpt_300-330.png}
        \caption{} 
        \label{fig:multi_front_centered_b}
    \end{subfigure}
    \begin{subfigure}[b]{0.495\linewidth}
        \centering
            \includegraphics[height=6cm, trim=0cm 0cm 0cm 0cm, clip]{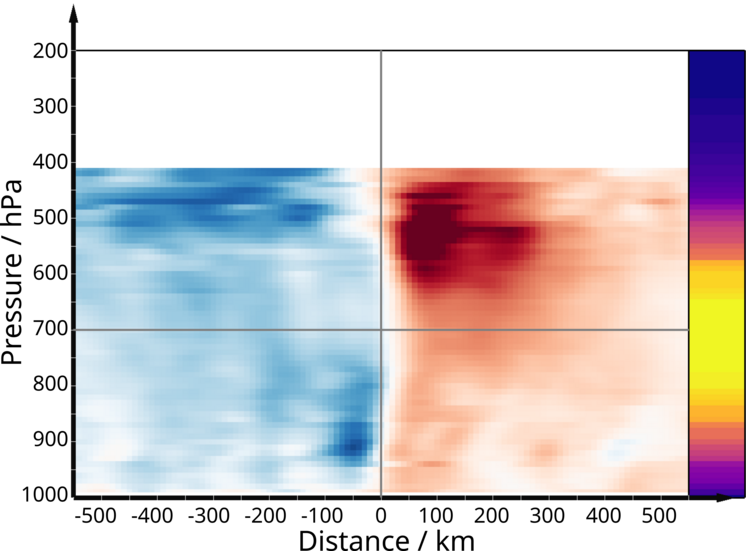} 
    \vspace{0.3cm}
    \centering
        \hfill
        \includegraphics[width=0.9\linewidth]{Colormap_vertVel_50_fronts.png}
        \caption{} 
        \label{fig:multi_front_centered_c}
    \end{subfigure}
    \hfill
    \begin{subfigure}[b]{0.495\linewidth}
        \centering
            \includegraphics[height=6cm,, trim=0.35cm 0cm 0cm 0cm, clip]{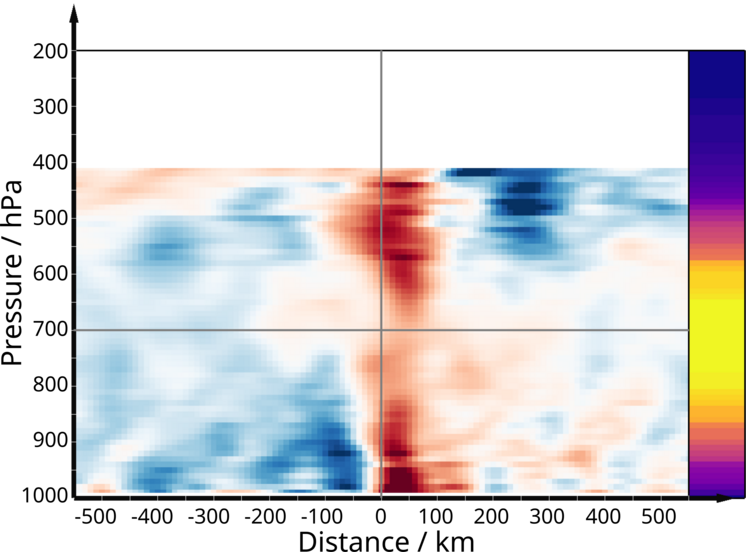} 
    \vspace{0.3cm}
    \centering
    \includegraphics[width=0.9\linewidth]{Colormap_convergence_50_fronts.png}
        \caption{} 
        \label{fig:multi_front_centered_d}
    \end{subfigure}
    \end{minipage}%
    \begin{minipage}[c]{0.06\linewidth}
        \centering
        \includegraphics[angle=90, height=6cm]{Colormap_dataAvailability_vertical.png}
    \end{minipage}

    \caption{Multiple-front \emph{centered all-pressure-levels mean}, that is, with the frontal position aligned to $u=0\,km$ at all levels for (a) horizontal gradient magnitude of $\theta_e$, (b) $\theta_e$, (c) vertical velocity (positive values indicate upward movement), and (d) horizontal convergence.} 

    \label{fig:multi_front_centered}
\end{figure}

\section{Discussion}
\label{sec:discussion}

In this paper, we have proposed a new approach to generating 2D vertical composite representations of frontal variables.
The approach enables the computation of vertical composites for both single fronts along the frontal line and multiple fronts.
At the core of our approach, we compute 2D front surfaces as contiguous structures from the 3D scalar field of the horizontal gradient magnitude of the equivalent potential temperature.
In contrast to previous methods that compute the fronts in 2D horizontal cross-sections only, we compute the front at all pressure levels simultaneously.
The extracted contiguous 2D front surfaces enable us to compute composite variables in such a way that patterns of the fronts and their close vicinity become more clearly visible.
However, the purpose of this work was not to perform a full analysis of frontal environments, but rather to provide a proof of concept that the presented approach enables a better analysis of near-front patterns and structures.
In order to exploit the full potential of the proposed approach, some limitations need to be addressed in future work.
Therefore, in the following, we discuss the current limitations as well as the preliminary results presented.

To extract the 2D front surface, we utilize a ridge surface extraction method~\cite{klenertLocalIterativeApproach2023}.
Although this method allows the extraction of front surfaces from unsmoothed data, it requires some initial seed points to start the computation.
For this, we used the 1D front lines computed in a previous study~\cite{paceyClimatologyNatureWarmseason2023}.
However, computation of the 2D front surface based on points sampled along these front lines failed in many cases because these front lines were computed on smoothed data.
As a result, in these cases, we had to resort to interactively correcting the seed points.
Furthermore, sometimes, when visually investigating the extracted fronts, we found that the fronts could be extended further.
Hence, additional seed points were manually placed at the boundaries of the fronts.
These manual interactions clearly limit the applicability of our approach, restricting it to either rather small sets of front cases or requiring a large investment of manual labor.
Both scenarios are not ideal.
Therefore, to make the proposed approach a useful alternative to previous methods, the 2D front surface extraction needs to be automated.

Using the extracted 2D front surfaces, we computed vertical composites of frontal variables by first extracting 2D vertical cross-sections that are horizontally orthogonal to the front line at a specific pressure level, here at 700\,hPa.
These vertical cross-sections were then normalized in two different ways: (1)~by placing the entire cross-section such that the front at 700\,hPa lies at the origin ($u=0\,km$), and (2)~by shifting each horizontal row of the cross-section such that the front positions at each pressure level are located at the origin.
With the latter novel approach, we transform the data such that the front line in each cross-section becomes a vertical line.
As a result, we can compute composites by averaging over cross-sections of a single front and even several fronts such that frontal variables with the same distance to the front are averaged over (\emph{centered all-pressure-levels mean}).
We further use the front surfaces to compute the mean vertical front line of each individual front, which are subsequently averaged and used to map the \emph{centered all-pressure-levels mean} back onto a representative vertical front structure.
The resulting composite thus shows the means obtained from averaging values at equal distances to the front, while also reflecting the mean vertical front structure (\emph{all-pressure-levels mean}).
This is in contrast to the previous approach (\emph{single-pressure-level mean}), where, depending on the variation of the frontal slope within a single front or the variation of the morphologies of different fronts, values corresponding to different horizontal distances to the front are averaged, leading to smeared composites (e.g., Fig.~\ref{fig:case_study_var_fields_a}, \ref{fig:multi_front_var_fields_a}).

With our new approach, we see clear patterns in the composites.
For the equivalent potential temperature, e.g., the composites show a clear separation of two distinct air masses along the mean vertical front line (Fig.~\ref{fig:case_study_var_fields_b}, \ref{fig:multi_front_var_fields_b}).
For the kinematic fields, a region of deep ascent and lower-tropospheric convergence ahead of the front as well as near-surface descent and divergence behind the front emerge clearly.
This is in accordance with our current conceptual understanding of cold-frontal dynamics and can thus be considered as proof of concept that our approach gives the desired results and shows a much clearer picture than previous approaches. 

However, since the sample size (50) considered in this study is not very large, we should be careful with general conclusions.
Although we see an intensification of some signals in the composites, the spread between individual cases is large, and there is still some noise in the composites, in particular for the kinematic variables.
Hence, the interpretation of the observed patterns requires a much larger sample size.
A further limitation of our study is the restriction to fronts in a certain region (Central Europe) and at a specific time (12 noon).
Therefore, in future work, we will investigate whether similar observations can be made for other data as well.

We should note that the choice of pressure level for computing the orientation of the vertical slices will have some influence on the final result of the composite.
This is because the orientations of the horizontal front lines will slightly change from pressure level to pressure level.
We expect this influence to be negligible since we do not assume the horizontal front lines to vary greatly along the pressure levels but to keep their general shape.
However, we did not test this, but instead we leave this for future work.

One inherent limitation of our approach is that it requires the front to be present at ``all'' pressure levels.
This is because we can only correct for the front position if we can identify a front position.
If no front can be extracted for some pressure level, this pressure level cannot be included in the computation of the composite.
Hence, our method introduces a bias towards fronts that extend over many pressure levels.
In order to work around this inherent limitation, we could interpolate front positions if they are missing due to holes in the 2D front surface, or extrapolate the fronts at the borders, in particular in the vertical direction.
However, as this might also introduce some errors, in this work we instead exclude pressure levels from the composite computation if no front was identified at the respective pressure level.
If the composite for a level is based on too few data, we ignore the level altogether.
As a result, our composites do not span all pressure levels, but are limited to a much smaller vertical range than the original data.
In particular, they do not span to the lowest pressure level, that is, they do not reach the Earth's surface.
In this study, we used a threshold of 10\,\%, that is, our results are based on at least 10\,\% of the data.
One reason for this threshold is the low number of fronts (50) that we looked at.
Here, a threshold of 10\,\% means that at least five fronts need to contribute to a pressure level.
When including more fronts in our statistics, we may be able to lower the threshold of 10\,\%, which would result in more complete composites.

By computing and visualizing the data availability for each vertical composite, we provide a first measure of uncertainty that makes the user aware that the mean values computed may be based on few data only.
However, further uncertainty measures would be desirable.
For example, it would be interesting to determine and compute appropriate measures to indicate the statistical relevance of the mean computed at each location over vertical composites.
However, such a computation does not seem to be straightforward, for which reason we leave this for future work.

\section{Conclusions}

In this work, we have presented a novel approach to the computation of vertical composite representations of frontal environments.
We believe that this approach will help to obtain a deeper understanding of the dynamics of atmospheric fronts.
However, to fully exploit the benefits of this method, a full automation of the front extraction method is required to enable the evaluation of a much larger number of fronts.
For the front extraction method used in this work, automation mainly concerns the generation of seed points.
We currently explore several methods to improve on this, including the scale-space approach by Kindlmann et al.~\cite{kindlmann_sampling_2009}.
However, we may also resort to other automated front extraction methods, if they fulfill the requirements to generate 2D contiguous front surfaces, ideally on unsmoothed input data.
Working on unsmoothed data, as was done in this work, may open the way to new analysis possibilities of ERA5 reanalysis data as well as additional model data with higher spatial resolution.
The automation of the front surface computation will finally allow us to carry out large-scale statistical analyses of 3D frontal environments.

\section*{Acknowledgments}
This research has been funded by Deutsche Forschungsgemeinschaft (DFG, German Research Foundation) within project C06
``Multiscale structure of atmospheric vortices'' of CRC 1114 ``Scaling Cascades in Complex Systems''. 
It was further supported by the Alexander von Humboldt Foundation and the cluster of excellence MATH+.

\bibliographystyle{plain} 
\bibliography{References}

\appendix

\section{Notation}
\label{sec:notations}

\begin{table}[H]
\centering
\caption{List of symbols and parameters.}
\label{tab:notations}
\begin{tabular}{lll}
\hline
\textbf{Symbol} & \textbf{Meaning} & \textbf{Unit} \\
\hline
$\theta$ & Potential temperature & K\\
$\theta_e$ & Equivalent potential temperature & K\\
$\lambda$ & Longitude & °\\
$\varphi$ & Latitude & °\\
$\nabla_h$ & Horizontal gradient operator & km$^{-1}$ \\
$\|\nabla_h \theta_e\|$ & Horizontal gradient magnitude of the equivalent potential temperature & K km$^{-1}$ \\%
$\mathcal{E}_\text{HG}$ & Horizontal extent of the grid of the vertical slice & km\\
$\mathcal{F}_{\text{HL}}$ & Horizontal front line \\
$\mathcal{F}_{\text{HL}}(i)$ & $i$th point of the horizontal front line \\
$\mathcal{F}_\text{S}$ & Front surface \\
$\mathcal{F}_{\text{VL}}$ & Vertical front line \\
$\mathcal{F}_{\text{VL}}(i)$ & Vertical front line at the $i$th horizontal front line point $\mathcal{F}_{\text{HL}}(i)$ \\

$G$ & Grid of the vertical slice \\

$\mathcal{L_\text{p}}$ & Pressure level & hPa\\

$\mathcal{N_{\text{HL}}}(i)$ & Normal vector of the horizontal front line at point $\mathcal{F}_{\text{HL}}(i)$\\

$\mathcal{P_{\text{H}\mathcal{L_\text{p}}}}$ & Horizontal plane at pressure level $\mathcal{L_\text{p}}$ \\
$\mathcal{P_{\text{V}}}(i)$ & Vertical plane at the $i$th horizontal front line point $\mathcal{F}_{\text{HL}}(i)$\\

$R$ & Radius of the Earth & km\\

$\mathcal{S_\text{HG}}$ & Horizontal spacing of the grid of the vertical slice & km\\
$\mathcal{S}_\text{HL}$ & Spacing of the horizontal front line & km\\
$\mathcal{S}_\text{VG}$ & Vertical spacing of the grid of the vertical slice & km\\
$\mathcal{S}_\text{VL}$ & Spacing of the vertical front line & km\\

$u$ & Horizontal coordinate of the vertical slice & km\\ 
$v$ & Vertical coordinate of the vertical slice & hPa\\ 

$X$ & Metric $x-$coordinate & km\\
$Y$ & Metric $y-$coordinate & km\\

\hline
\end{tabular}
\end{table}

\section{Cold Front Cases}
\label{sec:fronts}

\begin{table}[H]
    \centering
    \caption{Cold fronts used in this study. Dates and vertical ranges of the corresponding front surfaces.}
    \label{tab:list_fronts}
    
    \begin{tabular}{cccc}
    \hline
    \textbf{Front ID} & \textbf{Date} & \textbf{Min. level (hPa)} & \textbf{Max. level (hPa)} \\
    \hline
    1  & 2007-05-25 & 410 & 890 \\
    2  & 2007-06-25 & 370 & 990 \\
    3  & 2007-07-08 & 540 & 980\\
    4  & 2007-07-09 & 520 & 860 \\
    5  & 2007-07-18 & 600 & 900 \\
    6  & 2007-08-06 & 560 & 930\\
    7  & 2007-08-24 & 510 & 930 \\
    8  & 2008-07-11 & 520 & 920 \\
    9  & 2008-07-12 & 530 & 800 \\
    10 & 2008-08-01 & 580 & 830\\
    11 & 2008-08-12 & 390 & 860 \\
    12 & 2008-08-19 & 500 & 850 \\
    13 & 2009-05-18 & 470 & 830\\
    14 & 2009-06-15 & 320 & 870 \\
    15 & 2009-06-16 & 580 & 890 \\
    16 & 2009-07-14 & 470 & 860 \\
    17 & 2009-09-20 & 450 & 950\\
    18 & 2010-07-12 & 650 & 800 \\
    19 & 2011-05-19 & 630 & 890 \\
    20 & 2011-05-31 & 560 & 830\\
    21 & 2011-06-16 & 590 & 870 \\
    22 & 2011-06-18 & 480 & 770 \\
    23 & 2011-07-18 & 640 & 920 \\
    24 & 2011-08-06 & 590 & 990\\
    25 & 2011-08-21 & 650 & 980 \\
    26 & 2011-08-22 & 460 & 900 \\
    27 & 2012-05-11 & 470 & 940\\
    28 & 2012-07-27 & 550 & 990 \\
    29 & 2012-08-20 & 580 & 900 \\
    30 & 2013-08-19 & 510 & 930 \\
    31 & 2013-09-07 & 580 & 910 \\
    32 & 2013-09-16 & 580 & 930\\
    33 & 2014-06-08 & 580 & 860 \\
    34 & 2014-06-10 & 490 & 870 \\
    35 & 2014-07-30 & 550 & 860\\
    36 & 2014-08-10 & 600 & 970 \\
    37 & 2014-08-13 & 500 & 860 \\
    38 & 2014-08-29 & 490 & 990 \\
    39 & 2015-05-20 & 510 & 990\\
    40 & 2015-07-29 & 580 & 980 \\
    41 & 2015-08-07 & 480 & 940 \\
    42 & 2015-08-12 & 290 & 970\\
    43 & 2015-08-25 & 570 & 820 \\
    44 & 2015-09-01 & 490 & 960 \\
    45 & 2016-05-03 & 580 & 980 \\
    46 & 2016-07-01 & 580 & 930\\
    47 & 2016-07-02 & 440 & 940 \\
    48 & 2016-07-11 & 430 & 810 \\
    49 & 2016-08-20 & 550 & 910\\
    50 & 2016-09-11 & 480 & 930 \\
    \hline
    \end{tabular}
\end{table}

\section{Composite Histograms}
\label{sec:compositehistograms}

\begin{figure}[H]
    \centering

    \begin{subfigure}[t]{0.49\textwidth}
        \centering
        \includegraphics[width=\linewidth]{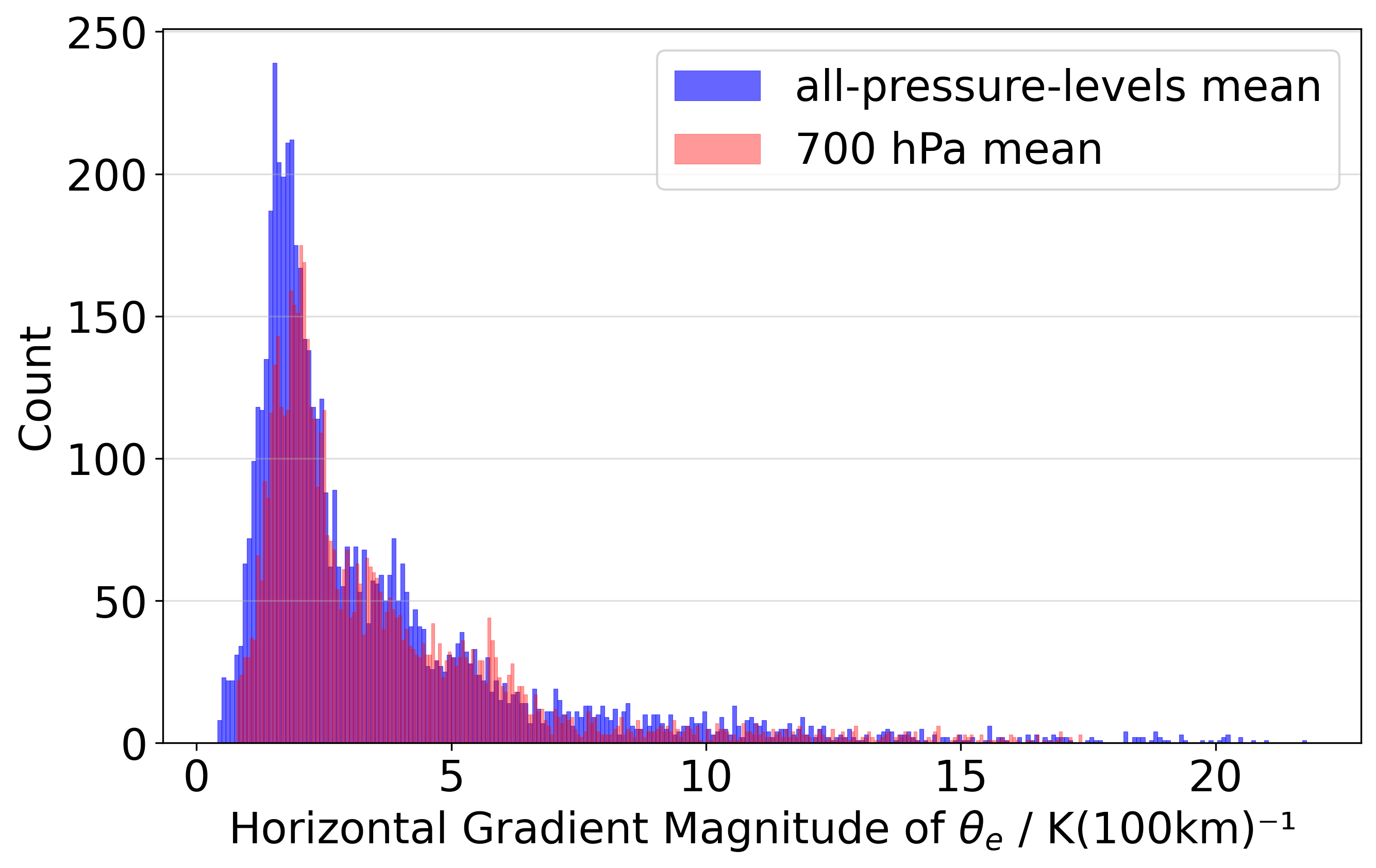}
        \caption{Single-front means for 11 May 2012}
    \end{subfigure}
    \hfill
    \begin{subfigure}[t]{0.49\textwidth}
        \centering
        \includegraphics[width=\linewidth]{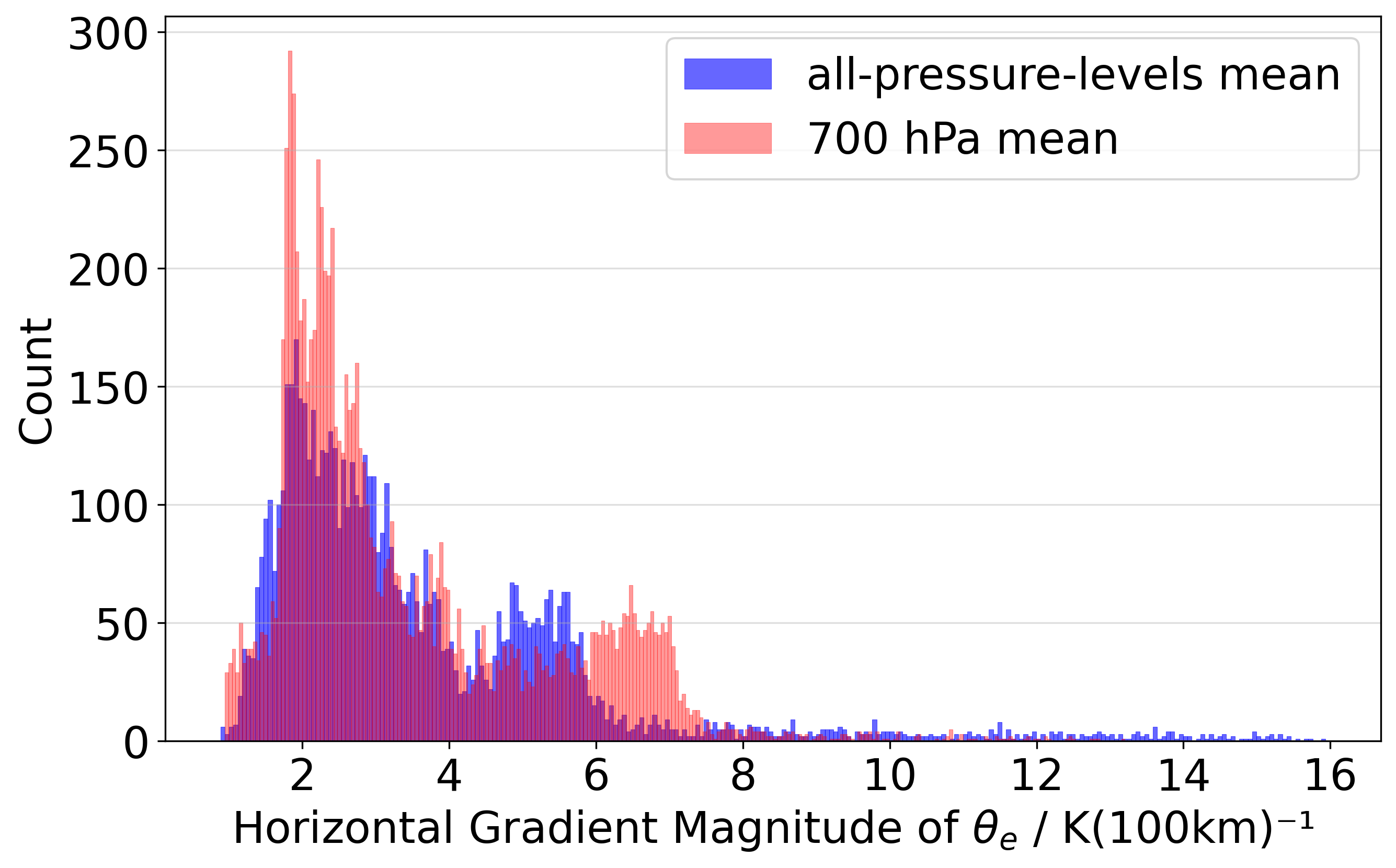}
        \caption{Multiple-fronts means}
    \end{subfigure}
    \caption{Histograms of the \emph{700\,hPa means} and \emph{all-pressure-levels means} of the horizontal gradient magnitude of $\theta_e$ for (a) the single-front composites presented in Sect.~\ref{sec:results:casestudy} (Fig.~\ref{fig:case_study_gradient}), and (b) the multiple-fronts composites presented in Sect.~\ref{sec:results:multiplefronts} (Fig.~\ref{fig:multi_front_gradient}).
    To ensure comparability, the histograms of the \emph{700\,hPa means} were calculated over the same domain as those of the \emph{all-pressure-levels means}. 
    The bin width was calculated by dividing the range of values of the respective variable by the number of bins (256).
    All histograms show that the majority of gradient values are small, while only a few values are high.
    Since the highest gradient values in the composites are associated with the frontal zone, this distribution reflects its narrow width relative to the frontal environment.
    Notably, the maximum values in the \emph{all-pressure-levels means} are approximately 20\% higher than those in the corresponding \emph{700\,hPa means}.
    In (b), the \emph{700\,hPa mean} contains a considerably larger number of values up to approximately 7 $\text{K}(100\text{ km})^{-1}$, than the \emph{all-pressure-levels mean}.
    This probably reflects the variability in frontal slopes, which leads to a reduction in the composite mean of the horizontal gradient magnitude of $\theta_e$.}
    \label{fig:histograms_gradient}
\end{figure}

\begin{figure}[H]
    \centering

    \begin{subfigure}[t]{0.49\textwidth}
        \centering
        \includegraphics[width=\linewidth]{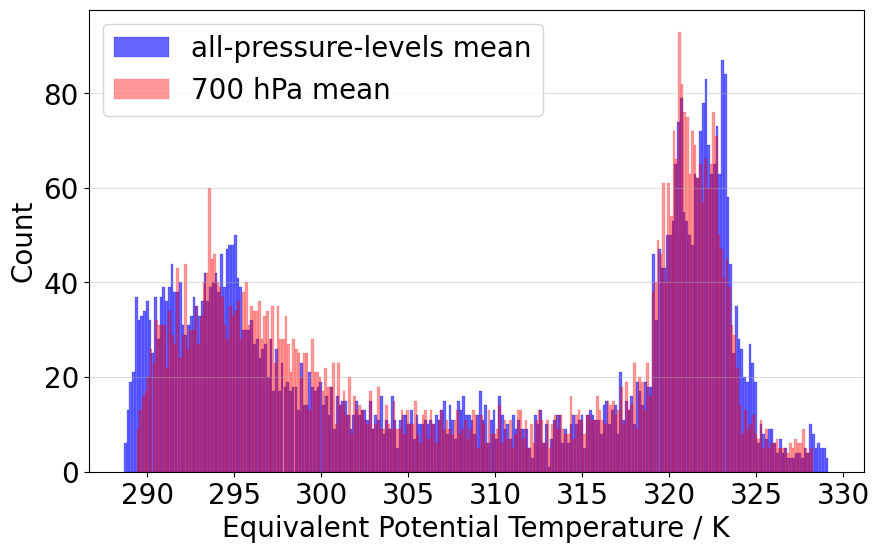}
        \includegraphics[width=\linewidth]{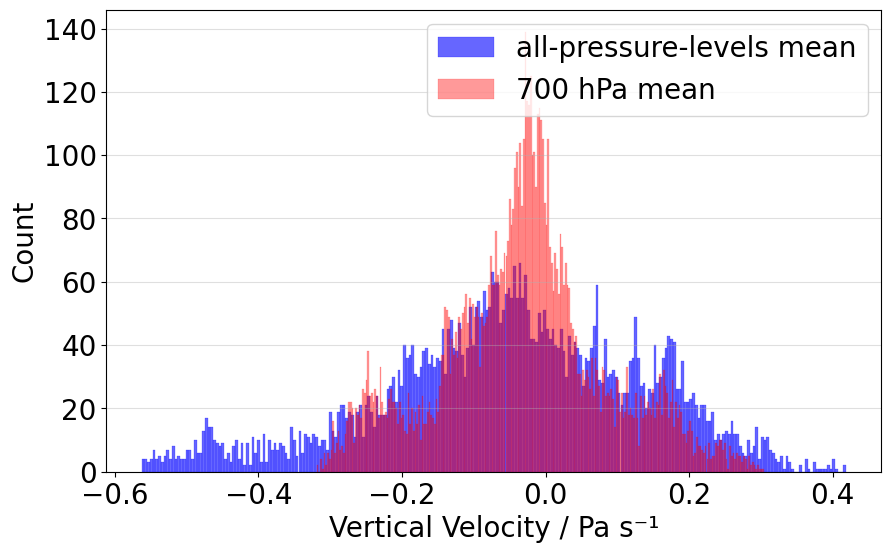}
        \includegraphics[width=\linewidth]{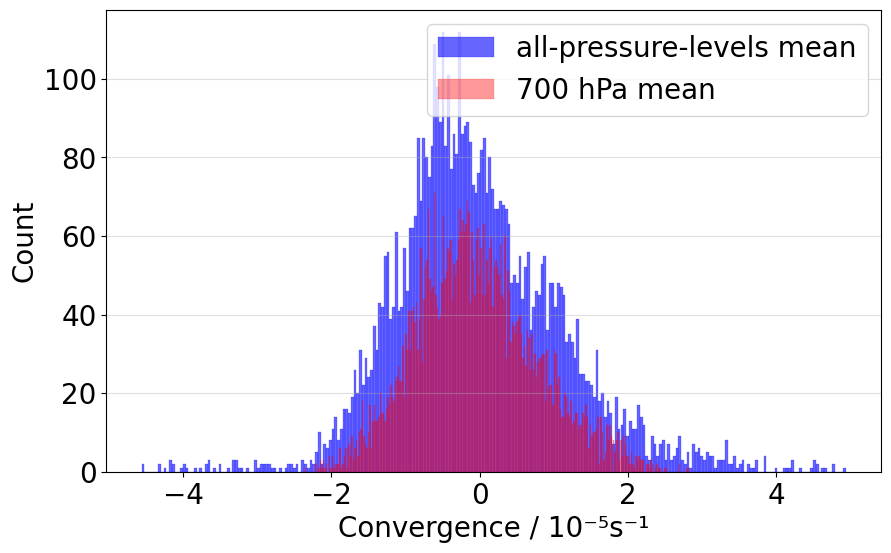}
        \caption{Single-front means for 11 May 2012}
    \end{subfigure}
    \hfill
    \begin{subfigure}[t]{0.49\textwidth}
        \centering
        \includegraphics[width=\linewidth]{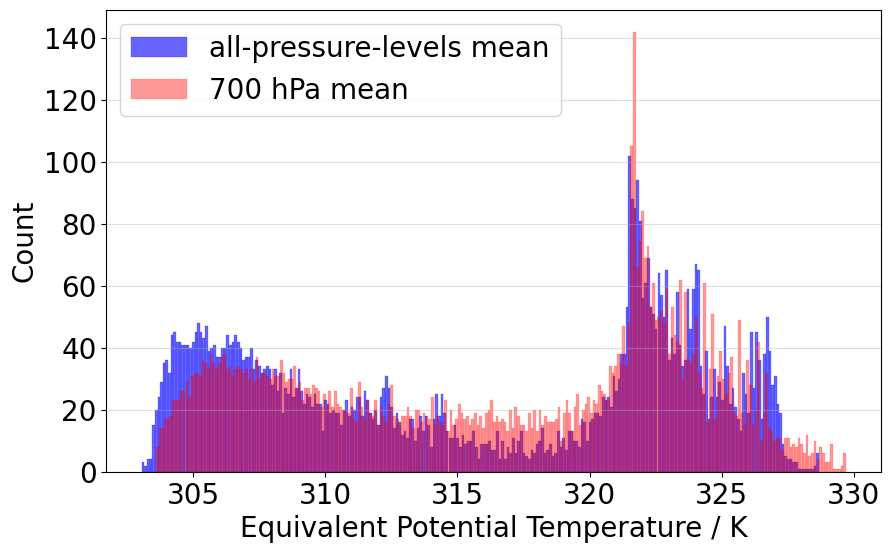}
        \includegraphics[width=\linewidth]{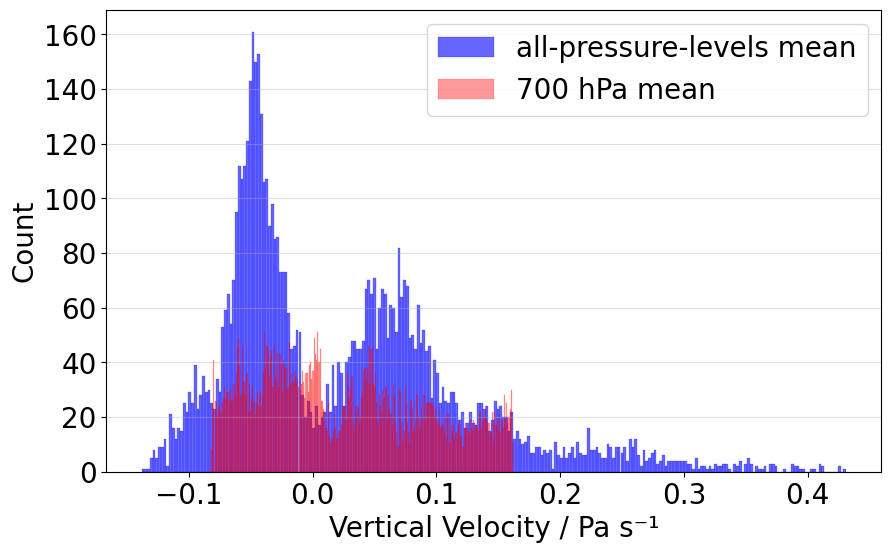}
        \includegraphics[width=\linewidth]{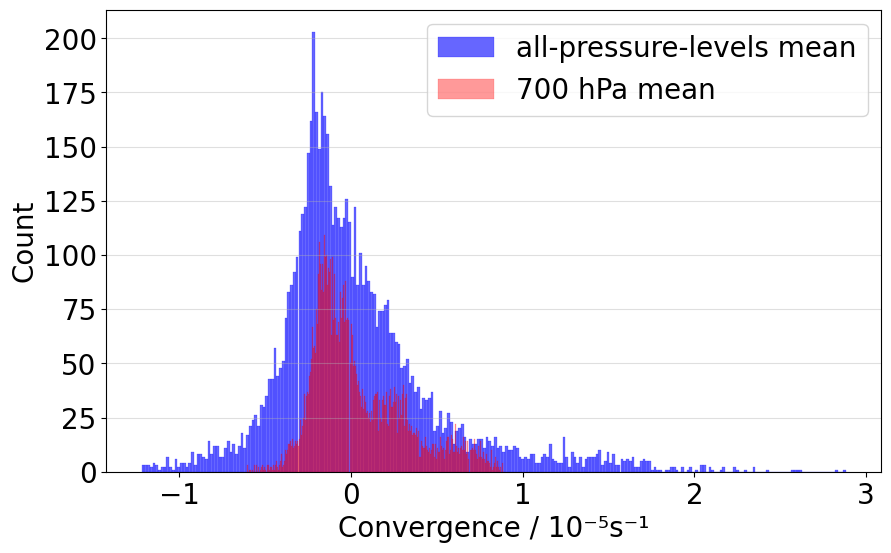}
        \caption{Multiple-fronts means}
    \end{subfigure}
    \caption{Same as Fig.~\ref{fig:histograms_gradient} but for $\theta_e$, vertical velocity and horizontal convergence composites of (a) the single front case presented in Sect.~\ref{sec:results:casestudy} (Fig.~\ref{fig:case_study_var_fields}), and (b) the 50 fronts presented in Sect.~\ref{sec:results:multiplefronts} (Fig.~\ref{fig:multi_front_var_fields}).
    The bimodal distribution of $\theta_e$-values reflects the presence of two different air masses separated by the front. 
    In particular, for the multiple-front \emph{all-pressure-levels mean}, the bimodal distribution shows a particularly deep minimum between the two modes, indicating a clearer separation between the two air masses.
    For vertical velocity and convergence, the values exhibit substantially larger magnitudes in the \emph{all-pressure-levels means} than in the corresponding \emph{700\,hPa means}.
    The histograms of vertical velocity further confirm the observations from Sect.~\ref{sec:results}: most values in the single-front \emph{700\,hPa mean} are close to zero, whereas the multiple-front \emph{all-pressure-levels mean} exhibits a bimodal distribution with a minimum near zero between the two modes, reflecting a clearer separation between upward and downward motion across the front.}
    
    \label{fig:histograms_vars}
\end{figure}

\end{document}